\documentclass[12pt]{iopart}

\usepackage{iopams} 
\usepackage{citesort}
\usepackage{xcolor}
\usepackage{graphicx}
\usepackage{ulem}

\eqnobysec

\begin{document}

\review[Scalar Dark Energy Models and Scalar-Tensor Gravity]{Scalar Dark Energy Models and Scalar-Tensor Gravity: Theoretical Explanations for the Accelerated Expansion of Present Universe}

\author{Peixiang Ji$^{1,2}$ and Lijing Shao$^{2,3}$}

\address{$^1$ Department of Astronomy, School of Physics, Peking University, Beijing 100871, China}
\address{$^2$ Kavli Institute for Astronomy and Astrophysics, Peking University, Beijing 100871, China}
\address{$^3$ National Astronomical Observatories, Chinese Academy of Sciences, Beijing 100012, China}
\ead{lshao@pku.edu.cn}
\vspace{10pt}

\begin{abstract}
The reason for the present accelerated expansion of the Universe stands as one
of the most profound questions in the realm of science, with deep connections to
both cosmology and fundamental physics. From a cosmological point of view,
physical models aimed at elucidating the observed expansion can be categorized
into two major classes: dark energy and modified gravity. We review various
major approaches that employ a single scalar field to account for the
accelerating phase of our present Universe.  Dynamical system analysis is
employed in several important models to seek for cosmological solutions that
exhibit an accelerating phase as an attractor.  For scalar field models of dark
energy, we consistently focus on addressing challenges related to the
fine-tuning and coincidence problems in cosmology, as well as exploring
potential solutions to them. For scalar-tensor theories and their
generalizations, we emphasize the importance of constraints on theoretical
parameters to ensure overall consistency with experimental tests.  Models or
theories that could potentially explain the Hubble tension are also emphasized
throughout this review.
\end{abstract}

%
%
%
%
%

\submitto{\CTP}

\section{Introduction}

In 1998, two independent research groups
\cite{SupernovaSearchTeam:1998fmf,SupernovaCosmologyProject:1998vns} studying
distant Type Ia Supernovae (SN Ia) presented compelling evidence that the
expansion of the Universe is not just expanding but actually accelerating.
Subsequent observations, which included more in-depth studies of supernovae
\cite{Podariu:1999ph,Riess:2006fw} and independent data from the cosmic microwave background
(CMB) radiation \cite{WMAP:2003elm,WMAP:2008ydk,WMAP:2008lyn,WMAP:2010qai,Planck:2013pxb,Planck:2015fie,Planck:2018vyg}, the baryon acoustic oscillation (BAO) peak
length scale \cite{SDSS:2005xqv,Blake:2011en,Beutler:2011hx,BOSS:2013rlg,Ross:2014qpa,BOSS:2016wmc,Ata:2017dya}, clusters of galaxies
\cite{Allen:2004cd,Chen:2004nqb}, and the large-scale structure (LSS) of the
Universe \cite{SDSS:2003eyi,SDSS:2004kqt,Percival:2007yw,DES:2016qvw}, confirmed this remarkable
acceleration.

According to Einstein's general theory of relativity without a cosmological
constant, if the Universe was primarily filled with ordinary matter and/or
radiation, which are the known constituents of the Universe, gravitational
interactions would act to slow down the expansion. However, since the Universe
is instead found to be accelerating, we are confronted with two intriguing
possibilities, dark energy and modified gravity, both of which would have
profound implications for our understanding of the Universe and the laws of
physics.

The first possibility is that approximately 70\% of the energy density of the
Universe exists in a new form characterized by negative pressure, referred to as
dark energy. When attributing it to the matter sector, dark energy would modify
the stress-energy tensor of matters in Einstein's field equations, leading to
the observed acceleration. This novel dark energy component has become central
to cosmology, and understanding its nature remains one of the most significant
challenges in modern cosmology and fundamental physics.

The second possibility is that general relativity, as formulated by Einstein,
breaks down when applied to cosmological scales. In this scenario, a more
comprehensive theory, modifying the Einstein-Hilbert action, would be needed to
accurately describe the behavior of the Universe, which would lead to changes in the
gravitational sector of the field equations.

Extra scalar fields besides the known Higgs field in the Standard Model of
particle physics may play a critical role in fundamental physics and provide
valuable insights into various aspects of the Universe~\cite{Will:2014kxa}.
Scalar fields naturally arise in unified theories, e.g. the string theory, and
contribute to the overall cosmic energy density, offering a plausible
explanation for the observed acceleration of the Universe
\cite{Padmanabhan:2002ji,Copeland:2006wr}. In the realm of gravitational
theories, scalar fields are pivotal in various models of modified gravity, such
as in scalar-tensor theories \cite{Damour:1992we,Faraoni:2004pi,Fujii:2003pa,Quiros:2019ktw}. The scalar field, in addition to the metric, describes
gravitational interactions, resulting in predictions that can deviate from those
of Einstein's theory of relativity. These deviations may manifest as changes in
the behavior of the gravitational force at different scales and under extreme
conditions, like those found near black holes, inside neutron stars, or during
the early Universe~\cite{Barausse:2020rsu,LISA:2022kgy,Doneva:2022ewd,Shao:2016ezh,Shao:2017gwu}.

The mystery of cosmic acceleration is connected to several pivotal questions in
cosmology and fundamental physics~\cite{Clifton:2011jh}. The cosmic acceleration
may hold the key to uncovering a successor to Einstein's theory of gravity. The
relatively small energy density of the quantum vacuum could yield insights into
concepts such as supersymmetry and superstring. The mechanism governing the
ongoing cosmic acceleration might relate to the primordial inflation in
cosmology, while the quest to unravel the cause of cosmic acceleration could
potentially introduce novel long-range forces or shed light on the enigmatic
smallness of neutrino masses in particle physics.

This review basically focuses on models and theories involving a single scalar field. The aim of this review is to theoretically clarify how these models drive the present Universe to undergo an accelerated expansion, with a wide utilization of dynamical system analysis. We typically do not delve into whether these models conform to various observations from SN Ia, CMB and LSS since it usually requires perturbation calculations and further analysis. For such discussions, the readers are directed to specific articles or reviews on observational constraints, such as Refs.~\cite{Peebles:2002gy,Avsajanishvili:2023jcl}.
In \Sref{CC}, we provide an
overview of the historical context of the cosmological constant, tracing its
birth from Einstein to its ultimate confirmation with supernova discoveries. The
cosmological constant problem is also overviewed in this section. Sections
\ref{DE} and \ref{Sec4} introduce scalar field models of dark energy and
modified gravity theories incorporating a single scalar field, respectively, which
offer explanations for the current expansion of the Universe. 
We list a series of scalar dark energy models in \Sref{DE} to show that it is challenging for them to solve fine-tuning and coincidence problems simultaneously without getting into other troubles. Interacting dark energy models, which have received much attention recently, are also introduced in this section.
In \Sref{Sec4}, we additionally provide constraints from terrestrial and Solar System tests of a general class of scalar-tensor gravity. Screening mechanisms are briefly included as well to illustrate how to maintain consistency with experimental results when there is a stronger coupling between the scalar field and matter.
A brief summary is provided in the final section.

Throughout the review, we adopt natural units $c=\hbar=1$, and have a metric
signature $(-,+,+,+)$. We denote the Planck mass as $m_\mathrm{Pl}=G^{-1/2} \simeq
1.22\times10^{19}\;\mathrm{GeV}$ and the reduced Planck mass as
$M_\mathrm{Pl}=(8\pi G)^{-1/2} \simeq 2.44\times10^{18}\;\mathrm{GeV}$, where $G$
is the Newton's gravitational constant. 
When the unit $\kappa\equiv8\pi G=1$ is employed, we will explain it in advance.

\section{Cosmological Constant}\label{CC}

The field equation in general relativity  with the cosmological constant
$\Lambda$ is
\begin{equation}\label{GR}
G_{\mu\nu}+\Lambda g_{\mu\nu}=\kappa T_
{\mu\nu}^\mathrm{m},
\end{equation}
where $G_{\mu\nu}$, $g_{\mu\nu}$ and $T_{\mu\nu}^\mathrm{m}$ are the Einstein tensor,
metric tensor, and energy-momentum tensor for matter in the standard model of
particle physics, respectively. The theory that adds a cosmological constant
term can be considered as the simplest dark energy model. Introducing a
cosmological constant to Einstein's field equations is a reasonable improvement
in the framework of general relativity. Since we all know that the metric itself
and the Einstein tensor are the only two tensors constructed from the metric and
its derivatives up to the second order, that possess vanishing covariant
divergences and lead to the local conservation of the energy-momentum tensor.

If one moves the cosmological constant to the right-hand side, \Eref{GR}
becomes
\begin{equation}
	G_{\mu\nu}=\kappa(T_{\mu\nu}^\mathrm{m}+T^\Lambda_{\mu\nu}),
\end{equation}
where
\begin{equation}
	T_{\mu\nu}^\Lambda=-\frac{\Lambda}{\kappa}g_{\mu\nu}
\end{equation}
represents the energy-momentum of a new component (dark energy) outside the
standard model of particle physics. Compared with the energy-momentum tensor of
a perfect fluid
\begin{equation}\label{Tpf}
T^\mathrm{pf}_{\mu\nu}=(\rho_\mathrm{pf}+{p_\mathrm{pf}})u_\mu
u_\nu+p_\mathrm{pf}g_{\mu\nu},
\end{equation}
one has
\begin{equation}\label{rhoL}
	\rho_\Lambda=-{p_\Lambda}=\frac{\Lambda}{\kappa},
\end{equation}
which indicates that this new component has a constant energy density and a
negative pressure. As a result, any form of matter with constant energy density
will behave as a cosmological constant (and vice versa), because they enter the
field equation with the same mathematical structure.

\subsection{A Brief History of Cosmological Constant}\label{CChis}

When the field equation of general relativity was proposed, most physicists,
including Einstein himself, believed that the Universe was in a static state.
However, the original field equation was not able to describe a non-evolving
Universe. To address this, Einstein introduced a new constant of nature and
referred to it as the cosmological constant \cite{Einstein:1917ce}.
Incorporating this constant, Einstein obtained his static Universe,  which he at
first failed to notice that it is unstable \cite{Eddington:1930zz}. Einstein
also believed that the existence of the cosmological constant does not result in
any difference on small scales, such as within the Solar system, as long as it
is sufficiently small.

In 1929, Hubble made the groundbreaking discovery that distant galaxies
are indeed moving away from us, and he presented strong observational evidence
for the expansion of the Universe \cite{Hubble:1929ig}. It indicated that the
Universe is dynamic rather than static. Hence, concerning the original
motivation, the cosmological constant is no longer needed. 

At the end of the last century, the cosmological constant was reintroduced
several times to explain problems related to the relatively young Universe age
\cite{Sandage:1961zz}, as well as the peaking number counts of quasars at $z\sim2$
\cite{Petrosian:1967qh}, despite it is well-known that the peak has an
explanation of astrophysical origin rather than cosmological nowadays. A non-zero cosmological constant was also invoked in
order to reconcile the flat Universe predicted by inflation and some other estimations from observations \cite{Turner:1984nf,Peebles:1984ge}.
 
The final confirmation of a non-zero value of the cosmological constant is from
the accurate measurement of the luminosity distance of SN Ia as a function of
redshift \cite{SupernovaSearchTeam:1998fmf,SupernovaCosmologyProject:1998vns}.
To fit the results in the framework of general relativity, the simplest way is
to assume that there exists a new cosmic component, besides radiation and matter
(including both ordinary matter and dark matter). The new component has a
negative pressure, and dominates the Universe's present energy budget.

\subsection{The Cosmological Constant Problem}\label{CCpr}

Although the confrontation of \Eref{GR} with the Solar system and
galactic observations have already given an upper bound of the cosmological constant, a
tighter constraint comes from larger scale observations
\cite{Weinberg:1988cp,Carroll:1991mt} that $\Lambda$ is of the same order of the
present value of the Hubble parameter $H_0$. The corresponding energy density is
calculated directly from \Eref{rhoL},
\begin{equation}\label{eq:rhoLambda}
    \rho_\Lambda\approx\frac{H^2_0}{8\pi G}\sim10^{-47}\,\mathrm{GeV}^4.
\end{equation}

Theoretically, the constant energy density associated with quantum vacuum energy
is regarded as the source of the cosmological constant. Similar to the ground
state of a quantum harmonic oscillator, which possesses a non-zero zero-point
energy proportional to the oscillation frequency of its classical counterpart,
vacuum-to-vacuum fluctuations arise from the collective contributions of an
infinite number of oscillators capable of vibrating at all possible frequencies.
A cutoff frequency is required to prevent the divergence of vacuum energy
density. The cutoff wavenumber $k_\mathrm{c}=m_{\mathrm{Pl}}$ is considered generally,
for the belief of the validity of the quantum field theory and general
relativity below the Planck scale. Then vacuum energy density has an extremely
large value
\begin{equation}\label{eq:rhoVAC}
\rho_{\mathrm{vac}}\approx\frac{k^4_\mathrm{c}}{16\pi^2}\sim10^{74}\,\mathrm{GeV}^4,
\end{equation}
which is more than 120 orders of magnitude larger than the observational bound
of $\rho_\Lambda$. This huge discrepancy is known as the cosmological constant
problem, or the vacuum catastrophe.

Hints that the cosmological constant might be non-zero (see \Sref{CChis}) led Zel'dovich to
consider the cosmological constant in terms of vacuum energy in 1967
\cite{Zeldovich:1968ehl}, and since then, numerous solutions have been proposed
to address this mismatch of energy density \cite{Weinberg:1988cp,Rugh:2000ji,Martin:2012bt,Burgess:2013ara}. Apart from anthropic principles \cite{Weinberg:1988cp,Weinberg:1996xe},
the solutions can be broadly categorized into three groups. The first one is to
establish a unified framework for quantum gravity, specifically in exploring the
interplay between quantum vacuum and gravity \cite{Green:1987sp,Rovelli:2004tv,Rovelli:2014ssa}. The second category is to
``eliminate'' the vacuum energy. In the context of supersymmetry---an extension
of the standard model---the vacuum energy is exactly zero due to the equal
number of fermionic and bosonic degrees of freedom \cite{Wess:1992cp}. In addition, an alternative
approach, called Schwinger's source theory \cite{Schwinger:1967rg,Schwinger:1968rh,Schwinger:1968rq}, changes the interpretation of the
quantum field theory formalism, and the vacuum energy is no longer needed. The
last is to modify the gravity \cite{Clifton:2011jh,CANTATA:2021ktz}, aiming to acquire an effective cosmological
constant, which is actrually not an exact resolution of the vacuum catastrophe. However,
the modified gravity seems to become more crucial because of the discovery of a
tiny but non-zero cosmological constant. If a modified theory of gravity
accurately describes the evolution of the Universe and well explains the observations, it is plausible that at
least one of the following two statements holds.  (I) The vacuum energy of the
Universe is either zero or extremely small. It is possible that the quantum
zero-point energy of each field is offset by its higher order corrections, or
there exists a cancellation mechanism---maybe a hidden symmetry---of vacuum
energy among different fields. In such a scenario, one would need to further
scrutinize the explanations of experimental evidence for vacuum energy like the
Casimir effect, and the vacuum concept associated with quantum chromodynamics
and the  spontaneous symmetry breaking in the electroweak force.  (II) The
vacuum energy, being a purely quantum concept, does not function as a source
within classical gravity theory that exerts influence on the geometry of
spacetime and other degrees of freedom in modified gravity theories, e.g. scalar
fields. In this scenario, the connection between quantum field theory and
gravity theory poses significant challenges, while the establishment of a
unified framework for describing all fundamental forces remains a central
pursuit in modern theoretical physics.

\section{Dark Energy}\label{DE}

In the framework of general relativity, the radiation and barotropic matter, as
the only contents of the Universe, cannot induce an accelerated expansion. The
concept of dark energy, termed by Turner \cite{Huterer:1998qv,Perlmutter:1999jt} as a potential
constituent of the Universe, propels the ongoing studies of the accelerated
expansion. In the standard Lambda Cold Dark Matter ($\Lambda$CDM) model
\cite{Peebles:2002gy}, the dark energy is characterized by the presence of a
cosmological constant term, which is not dynamical but exerts influence on the
evolution of the Universe. Despite being the standard model in cosmology, it is
confronted with various theoretical and observational challenges.
	\begin{itemize}
	\item	{\bf Fine-tuning problem:} The current energy density of the estimated
	cosmological constant in \Eref{eq:rhoLambda} is relatively small
	based on observations, while the theoretical value of the vacuum energy
	density in \Eref{eq:rhoVAC} is significantly larger. In other
	words, the observed value of energy density conflicts with the possible
	energy scales and requires fine-tuning if the cosmological constant
	originates from the vacuum energy density.
	\item	{\bf Coincidence problem:} The $\Lambda$CDM model suggests a comparable
	energy density ratio between the matter and the dark energy
	\cite{Planck:2018vyg}, implying that the expansion of the Universe
	transitioned from a decelerated to an accelerated phase at a redshift of
	approximately $z\sim0.67$ \cite{Copeland:2006wr}. The problem of why an
	accelerated expansion should occur now (actually not long ago) in the very
	long history of the Universe is called the coincidence problem.
	\item	{\bf Hubble tension and more:} Recent observations have revealed that
	within the framework of $\Lambda$CDM model, there exists certain tensions
	between observations from early Universe and late Universe
	\cite{Verde:2019ivm}, among which the $H_0$ and $S_8$ tensions are the
	most well-known ones. The latest observations with the James Webb Space Telescope further confirm the existence of Hubble tension at $8\sigma$ confidence \cite{Riess:2024ohe}.
	\end{itemize}

\subsection{Quintessence}

A dynamical scalar field as dark energy---the quintessence model---is one of the most popular models for the dark energy. The audiences may be familiar with the similar inflation model that describes the very early stage of the Universe \cite{Guth:1980zm,Linde:1981mu}. However, except for the huge discrepancy in the energy scale, the scalar field is the only component in inflation models, while it is just the fifth element of the Universe in the quintessence model. As the first and the simplest model in this review, we will look at the quintessence in more detail. The evolution of the Universe in this model is displayed in \Sref{secq}, and the introduction of the classification and dynamical system analysis of the quintessence model are in \Sref{tfm} and \Sref{ADS}, respectively.

The action for the
quintessence model is given by \cite{Caldwell:1997ii}
\begin{equation}\label{SQ}
S[g^{\mu\nu},\phi,\Psi_\mathrm{m}]=S_\mathrm{EH}[g^{\mu\nu}] +
S_\mathrm{q}[g^{\mu\nu},\phi]+S_\mathrm{m}[g^{\mu\nu},\Psi_\mathrm{m}],
\end{equation}
where $S_\mathrm{EH}=\int \rmd^4x\sqrt{-g}R/2\kappa$ is the Einstein-Hilbert
action, $S_\mathrm{m}=\int \rmd^4x\sqrt{-g}\mathcal L_\mathrm{m}(\Psi_\mathrm{m})$ is the action of all
possible matter fields, collectively denoted as $\Psi_\mathrm{m}$, and
\begin{equation}\label{QA}
	S_\mathrm{q}=\int \rmd^4x\sqrt{-g}\mathcal L_\mathrm{q}=\int
	\rmd^4x\sqrt{-g}\left[-\frac12g^{\mu\nu}
	\partial_\mu\phi\partial_\nu\phi-V(\phi)\right],
\end{equation}
is the action for the quintessence. $V(\phi)$ is the potential of the scalar field and is always assumed to be positive. By variating the action \eref{SQ},
equations of motion for the tensor field $g_{\mu\nu}$ and the scalar field
$\phi$ are
\begin{eqnarray}
G_{\mu\nu}=\kappa\big(T_{\mu\nu}^\mathrm{m}+T_{\mu\nu}^\mathrm{q}\big),\label{GRQ}\\
	\square\phi =\frac{dV}{d\phi}\label{eqphi}, 
\end{eqnarray}
where $\square=g^{\mu\nu}\nabla_\mu\nabla_\nu$, and $T_{\mu\nu}^\mathrm{m}$ is the
energy-momentum tensor of ordinary matter including dust and radiation,
\begin{equation}\label{T}
T_{\mu\nu}^\mathrm{m}\equiv-\frac{2}{\sqrt{-g}}
\frac{\delta(\sqrt{-g}\mathcal{L}_\mathrm{m})}{\delta g^{\mu\nu}},
\end{equation}
and the energy-momentum tensor of quintessence is
\begin{equation}\label{TQ}
T_{\mu\nu}^\mathrm{q}=\partial_\mu\phi\partial_\nu\phi -
g_{\mu\nu}\left[\frac12g^{\alpha\beta}
\partial_\alpha\phi\partial_\beta\phi+V(\phi)\right].
\end{equation}

Since the scalar field is minimally coupled to the metric, the action \eref{SQ}
can be rewritten as
\begin{equation}\label{SQ'}
	S[g^{\mu\nu},\phi,\Psi_\mathrm{m}]=S_\mathrm{EH}
	[g^{\mu\nu}]+S'_\mathrm{m}[g^{\mu\nu},\Psi_\mathrm{m},\phi],
\end{equation}
where $S_\mathrm{m}'=S_\mathrm{m}+S_\mathrm{q}$.  In such a case, the scalar field in action \eref{SQ'}
represents the fifth element of the Universe, other than baryons, photons,
neutrinos and dark matter that are usually considered in modern cosmology. This
dynamical scalar field as a component of the Universe has first been studied by
Ratra \& Peebles \cite{Ratra:1987rm} and Wetterich \cite{Wetterich:1987fm} in
1988, and was first called quintessence in a 1998 paper by Caldwell et al.
\cite{Caldwell:1997ii}.

\subsubsection{Evolution of the Universe in the Quintessence Model.}\label{secq}
Based upon the assumptions of homogeneity and isotropy of the Universe, which is
approximately true on large scales, the Friedmann-Lema\^{i}tre-Robertson-Walker
(FLRW) metric,
\begin{equation}
	\rmd s^2=-\rmd t^2+a^2(t)\left(\frac{\rmd r^2}{1-Kr^2}+r^2\rmd\Omega^2_2\right),
\end{equation}
is used to describe the geometry of the Universe. Here, $a(t)$ is the scale
factor with the cosmic time $t$, and $\rmd\Omega_2^2=\rmd\theta^2+\sin^2\theta\rmd\phi^2$\footnote{Here $\phi$ is the azimuthal angle, not to be confused with the scalar field.} is the line element of a
2-sphere. The constant $K$ describes the geometry of the spatial part of the
spacetime, with closed, flat, and open Universes corresponding to $K = +1, 0,
-1$, respectively. In this review, we always assume a flat Universe, which is
consistent with current observations \cite{Planck:2018vyg}.

The Friedmann equations are a set of equations that govern the expansion of
space in homogeneous and isotropic models of the Universe. Using the FLRW
metric, one has
\begin{eqnarray}
	3M_\mathrm{Pl}^2H^2=\rho_\mathrm{m}+\rho_\mathrm{q},\label{QRW}\\
	-2M_\mathrm{Pl}^2\dot H=\rho_\mathrm{m}+\rho_\mathrm{q}+p_\mathrm{m}+p_\mathrm{q}\label{QRW'},
\end{eqnarray}
where $H=\dot a/a$ is the Hubble parameter. The subscripts ``$\mathrm{q}$" and ``$\mathrm{m}$" represent
quintessence and ordinary matter, respectively. Ordinary matter comprises relativistic radiation and non-relativistic dust, denoted by subscripts ``$\mathrm{r}$" and ``$\mathrm{d}$" respectively in the following. Assuming that the scalar field
only depends on the cosmic time, the energy density and pressure of the
quintessence field are
\begin{eqnarray}
\rho_\mathrm{q} =\frac12\dot\phi^2+V(\phi),\label{rhoQ}\\
p_\mathrm{q} =\frac12\dot\phi^2-V(\phi),\label{pQ}
\end{eqnarray}
where the overdot denotes the derivative with respect to the cosmic time $t$. Furthermore, local
conservation of the energy-momentum tensor gives
\begin{eqnarray}
	\dot\rho_\mathrm{m}+3H(1+w_\mathrm{m})\rho_\mathrm{m}=0,\label{Bm}\\
	\dot\rho_\mathrm{q}+3H(1+w_\mathrm{q})\rho_\mathrm{q}=0,\label{BQ}
\end{eqnarray}
where $w_\mathrm{m}=p_\mathrm{m}/\rho_\mathrm{m}$ and
\begin{equation}
w_\mathrm{q}=\frac{p_\mathrm{q}}{\rho_\mathrm{q}}=\frac{\dot\phi^2-2V(\phi)}{\dot\phi^2+2V(\phi)},
\end{equation}
are parameters in equations of state (EOS). In the slow-roll limit,
$\dot\phi^2\ll V(\phi)$, the parameter $w_\mathrm{q}$ goes to $-1$ and the quintessence
acts just like the cosmological constant. 

\begin{figure}[h]
    \centering
    \includegraphics[width=0.75\textwidth]{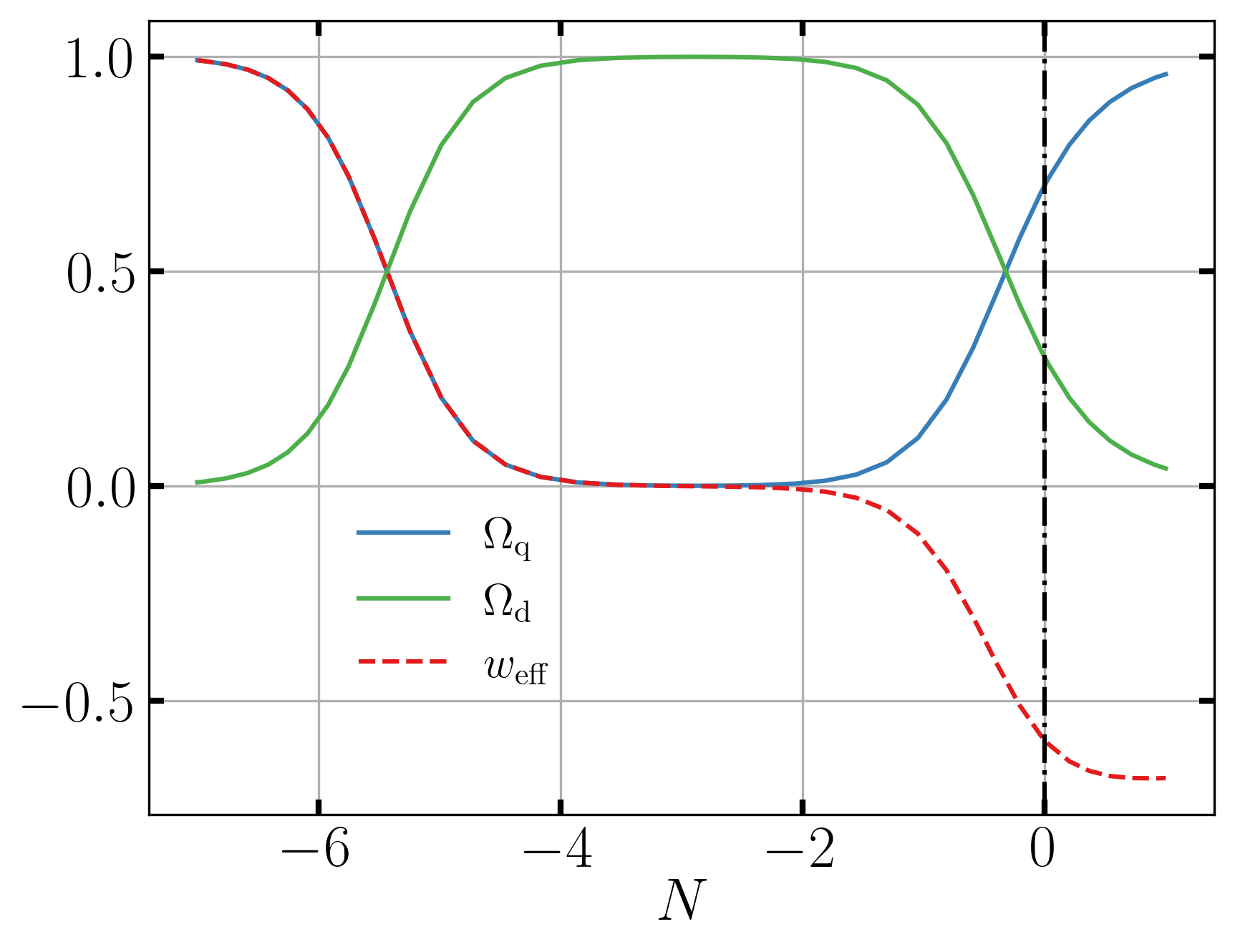}
    \caption{Evolution of the effective EOS parameter $w_\mathrm{eff}$, the dust $\Omega_\mathrm{d}$ and quintessence $\Omega_\mathrm{q}$ relative energy densities with an exponential potential. The vertical dashed line denotes the present cosmological time. One has $\Omega_\mathrm{d}\simeq0.3$ and $\Omega_\mathrm{q}\simeq0.7$ today.}
    \label{f:qO1}
\end{figure}

For a constant $w_\mathrm{m}$, \Eref{Bm} gives
\begin{equation}
	\rho_\mathrm{m}=\bar\rho_\mathrm{m} a^{-3(1+w_\mathrm{m})},
\end{equation}
while the solution to \Eref{BQ} is
\begin{equation}
\rho_\mathrm{q}=\bar\rho_{\mathrm{q}}\exp\left[-3\int(1+w_\mathrm{q})\frac{\rmd a}{a}\right].
\end{equation}
Here, $\bar\rho_\mathrm{m}$ and $\bar\rho_\mathrm{q}$ are integration constants. \Fref{f:qO1} depicts the evolution of dust and quintessence energy density parameters, as well as the effective EOS parameter of the quintessence model with an exponential potential, over time. The definitions of the quantities in the plot are provided in \Sref{ADS}. As observed from \Fref{f:qO1}, to satisfy $\Omega_\mathrm{d}\simeq0.3$ and $\Omega_\mathrm{q}\simeq0.7$ at present and to experience a period of matter dominance, the initial conditions must be chosen carefully.

We are currently in a period characterized by the dominance of dark energy. In
the quintessence domination, Equations \eref{QRW} and \eref{QRW'} become
\begin{eqnarray}
	6M_\mathrm{Pl}^2H^2=\dot\phi^2+2V(\phi),\label{QD}\\
	-2M_\mathrm{Pl}^2\dot H=\dot\phi^2\label{QD'}.
\end{eqnarray}
The acceleration equation comes out of a combination of the Friedmann equations,
as
\begin{equation}
-3M_\mathrm{Pl}^2\frac{\ddot a}{a}=\frac12(\rho_\mathrm{q}+3p_\mathrm{q})=\dot\phi^2-V(\phi),
\end{equation}
which shows that the Universe expands in an increasing rate when $w_\mathrm{q}<-1/3$,
or $\dot\phi^2<V(\phi)$. To see how the scalar field evolves, we assume a
power-law expansion, $a(t)\propto t^s$, where the accelerated expansion occurs
for $s>1$. According to \Eref{QD'}, one obtains
\begin{equation}\label{phit}
	\phi=\int \rmd t\sqrt{-2M_\mathrm{Pl}^2\dot H}\propto\ln t.
\end{equation}
The potential can also be expressed in terms of $H$ and $\dot H$ as
\begin{equation}\label{Vt}
	V=3M_\mathrm{Pl}^2H^2\left(1+\frac{\dot H}{3H^2}\right)\propto t^{-2}.
\end{equation}
Combining Equations \eref{phit} and \eref{Vt}, the potential giving the
power-law expansion is ought to be an exponential form
\begin{equation}\label{V}
V(\phi)=\bar V\exp\left(-\frac{\lambda\phi}{M_\mathrm{Pl}}\right),
\end{equation}
where $\bar V$ is a constant, and $\lambda=\sqrt{2/s}$.

\subsubsection{Thawing and Freezing Models.}\label{tfm}

In the flat FLRW background, \Eref{eqphi} gives the evolution equation
of the quintessence,
\begin{equation}\label{ball}
\ddot\phi+3H\dot\phi+\frac{\rmd V}{\rmd\phi}=0.
\end{equation}
Although the presence of ordinary matter is not explicitly accounted for in
\Eref{ball}, its influence on the Hubble parameter will impact the
evolution of the quintessence field. In \Eref{ball}, the first term
represents the acceleration of the scalar field, while the second term
corresponds to a frictional effect, known as Hubble friction or Hubble drag,
resulting from the expansion of the Universe, and the third denotes a driving
force arising from the steepness of the potential.

The potential function of the scalar field is the only adjustable component in
the quintessence model, whose specific shape exerts a significant influence on
the cosmic evolution. In order to reproduce the cosmic acceleration today we
require that the potential is flat enough to satisfy the condition
\begin{equation}
	\left|\frac{M^2_\mathrm{Pl}}{V}\frac{\rmd^2V}{\rmd\phi^2}\right|\lesssim1.
\end{equation}
Hence, the quintessence mass squared $m^2_\phi\equiv \rmd^2V/\rmd\phi^2$ needs to
satisfy
\begin{equation}
	|m^2_\phi|\lesssim\frac{V_0}{M^2_\mathrm{Pl}}\sim H_0^2,
\end{equation}
where $V_0$ and $H_0$ are the present values of the potential and Hubble
parameter, respectively. As a result, the mass of the scalar field has to be
extremely small, say $m_\phi\lesssim10^{-33}\;\mathrm{eV}$, to be compatible
with the present cosmic acceleration. Apart from this requirement, we also hope
that the potential appears in some particle physics models
\cite{Copeland:2006wr}.

Caldwell and Linder~\cite{Caldwell:2005tm} divided quintessence models into two
categories according to whether the scalar field accelerates ($\ddot\phi>0$) or
decelerates ($\ddot\phi<0$). A coasting in the scalar field dynamics,
$\ddot\phi=0$, is nongeneric, as the field would need to be finely tuned to be
perfectly balanced, neither accelerate due to the slope of the potential nor
decelerate due to the Hubble drag.

In the accelerating region, the field that has been frozen by the Hubble
friction into a cosmological-constant-like state until recently and then begins
to evolve gives the thawing model. 
The current accelerated expansion is driven by the dominance of quintessence and the scalar field in its early stages of ``thawing", implying that $w_q$, while increasing, remains smaller than $-1/3$.
A representative potential of this class is
the one arising from the pseudo Nambu-Goldstone boson \cite{Frieman:1995pm},
\begin{equation}
	V(\phi)=\mu^4\left[1+\cos\left(\frac{\phi}{f}\right)\right],
\end{equation}
where $\mu$ and $f$ are constants characterizing the energy scale and the mass
scale of spontaneous symmetry breaking, respectively. If the field mass squared
$|m^2_\phi|\simeq\mu^4/f^2$ around the potential maximum is smaller than $H^2$,
the field is stuck there with $w_\mathrm{q}\simeq-1$. Once $H^2$ drops below
$|m^2_\phi|$, the scalar field starts to move and $w_\mathrm{q}$ starts to increase.
Likelihood analysis using CMB shift parameters measured by WMAP7
\cite{WMAP:2010qai} combined with SN Ia and BAO data placed a bound, $w_\mathrm{q}|_{z =
0} < -0.695$, at the 95 \% confidence level, when the quintessence prior is
assumed to be $w_\mathrm{q}>-1$ \cite{Chiba:2012cb}.

In the decelerating region, the field that initially rolls due to the steepness
of the potential but later approaches the cosmological constant state gives the
freezing model. A representative potential of this class is the Ratra-Peebles
potential \cite{Ratra:1987rm,Zlatev:1998tr},
\begin{equation}\label{RPV}
	V(\phi)=M^{4+n}\phi^{-n},
\end{equation}
where $M$ and $n$ are positive constants. The potential does not possess a
minimum and hence the field rolls down the potential towards infinity. The
movement of the field gradually slows down when $\phi$ is large enough and the
system enters the phase of cosmic acceleration. Under the quintessence prior
$w_\mathrm{q}^{\rmi} > -1$ for tracking solutions (see \Sref{ADS}), likelihood
analysis using the iterative solution in Ref.~\cite{Chiba:2009gg} resulted in
bounds, $w_q^{\rmi} < -0.923$ and $0.675<\Omega_\mathrm{q}< 0.703$, at the 95 \% confidence
level, from the observational data of Planck 2015, SN Ia, and CMB
\cite{Durrive:2018quo}. For the Ratra-Peebles potential, this bound translates
to $n < 0.17$, which means that the tracker solutions arising from the inverse
power-law potential with positive integer powers are observationally excluded.

\subsubsection{Dynamical System Approach.}\label{ADS}

The Universe comprises multiple components, and each of these exerts its own
influence on the dynamic evolution of the Universe.
The dynamical system approach is a powerful method that helps illustrate the global dynamics and full evolution of the Universe \cite{Coley:2003mj,Bahamonde:2017ize}. Through the careful choice of the dynamical variables, a given cosmological model can be written as an autonomous system of some differential equations, which further provides a quick answer to whether it can reproduce the observed expansion of the Universe.

An autonomous system is a system of ordinary differential equations which does not explicitly depend on the independent variable,
$$\frac{\rmd}{\rmd t}\mathbf{x}(t)=f(\mathbf{x}(t)),$$
where $\mathbf{x}$ takes values in $n$-dimensional Euclidean space. Many laws in physics, where $t$ is often interpreted as time, are expressed as autonomous systems because it is always assumed that the laws of nature which hold now are identical to those for any point in the past or future. The time $t$ is usually replaced by the scale factor $a$ or its logarithm in cosmology because the Universe expands monotonously in most of cases so that the scale can be used as a measure of time.

In order to describe the quintessence model, the following dimensionless variables
\begin{equation}\label{ds}
x_\mathrm{K}\equiv\frac{\dot\phi}{\sqrt{6}M_\mathrm{Pl}H},\quad
x_\mathrm{V}\equiv\frac{\sqrt{V}}{\sqrt{3}M_\mathrm{Pl}H},\quad
x_\mathrm{r}\equiv\frac{\sqrt{\rho_\mathrm{r}}}{\sqrt{3}M_\mathrm{Pl}H},
\end{equation}
were introduced in the autonomous system \cite{Copeland:1997et,Tsujikawa:2010sc}. These above variables are generally called the expansion normalised variables \cite{2005dsc..book.....W} that are widely employed in quintessence models and their generalizations, for the sake of easily closing the dynamical system.
Furthermore, one defines density parameters for the kinetic term
and potential of the scalar field,
\begin{equation}
	\Omega_\mathrm{K}\equiv\frac{\dot\phi^2/2}{3M_\mathrm{Pl}^2H^2}=x_\mathrm{K}^2,\quad
	\Omega_\mathrm{V}\equiv\frac{V}{3M_\mathrm{Pl}^2H^2}=x_\mathrm{V}^2,
\end{equation}
which give
\begin{equation}\label{OQ}
	\Omega_\mathrm{q}\equiv\frac{\rho_\mathrm{q}}{3M_\mathrm{Pl}^2H^2}=x_\mathrm{K}^2+x_\mathrm{V}^2.
\end{equation}
Density parameters of radiation, dust, and the total ordinary fluid are given by
\begin{equation}
	\Omega_\mathrm{r}\equiv\frac{\rho_\mathrm{r}}{3M_\mathrm{Pl}^2H^2},\quad
	\Omega_\mathrm{d}\equiv\frac{\rho_\mathrm{d}}{3M_\mathrm{Pl}^2H^2},\quad
	\Omega_\mathrm{m}\equiv\frac{\rho_\mathrm{m}}{3M_\mathrm{Pl}^2H^2}=\Omega_\mathrm{r}+\Omega_\mathrm{d}.
\end{equation}
From \Eref{QRW}, one has
\begin{equation}\label{flat}
\Omega_\mathrm{q}+\Omega_\mathrm{m}=\Omega_\mathrm{K}+\Omega_\mathrm{V}+\Omega_\mathrm{d}+\Omega_\mathrm{r}=1,
\end{equation}
which implies the flatness of the Universe. The EOS parameter $w_\mathrm{q}$ and the
effective EOS parameter of the Universe are expressed by $x_\mathrm{K}$ and $x_\mathrm{V}$, via \cite{Tsujikawa:2010sc}
\begin{equation}\label{WQ}
w_\mathrm{q}=\frac{x_\mathrm{K}^2-x_\mathrm{V}^2}{x_\mathrm{K}^2+x_\mathrm{V}^2},\quad\quad
w_\mathrm{eff}\equiv\frac{p_\mathrm{m}+p_\mathrm{q}}{\rho_\mathrm{m}+\rho_\mathrm{q}}=x_\mathrm{K}^2-x_\mathrm{V}^2+\frac13x_\mathrm{r}^2.
\end{equation}

Differentiating the variables $x_\mathrm{K}$, $x_\mathrm{V}$ and $x_\mathrm{r}$ with respect to the number
of $\mathrm{e}$-foldings, $N=\ln a$, as well as using Equations \eref{QRW},
\eref{QRW'}, and \eref{ball}, one obtains the following first-order
differential equations \cite{Tsujikawa:2010sc}
\numparts
\begin{eqnarray}
\frac{{\rmd} x_\mathrm{K}}{\rmd N}=-3 x_\mathrm{K}+\frac{\sqrt{6}}{2} \lambda x_\mathrm{V}^2-x_\mathrm{K} \frac{1}{H}
\frac{\rmd H}{\rmd N},\label{DS1}\\
\frac{\rmd x_\mathrm{V}}{\rmd N}=-\frac{\sqrt{6}}{2} \lambda x_\mathrm{K} x_\mathrm{V}-x_\mathrm{V} \frac{1}{H} \frac{\rmd
H}{\rmd N},\label{DS2}\\
\frac{\rmd x_\mathrm{r}}{\rmd N}=-2 x_\mathrm{r}-x_\mathrm{r} \frac{1}{H} \frac{\rmd H}{\rmd N},\label{DS3}
\end{eqnarray}\label{DS123}
\endnumparts
where
\begin{equation}
	\frac{1}{H} \frac{\rmd H}{\rmd N}=\frac{\dot H}{H^2}=-\frac{1}{2}(3+3x_\mathrm{K}^2-3x^2_\mathrm{V}+x_\mathrm{r}^2),
\end{equation}
and $\lambda$ is defined by
\begin{equation}\label{lambda}
	\lambda\equiv-\frac{M_\mathrm{Pl}}{V}\frac{\rmd V}{\rmd\phi}.
\end{equation}

If $\lambda$ is constant, the differential equations are closed to become an autonomous system, and fortunately, the potential is exactly the form of \Eref{V}. In this way the analysis of the features of the phase space of this system, i.e. the analysis of the fixed points and the determination of the general behaviour of the orbits, provides insight on the global behaviour of the Universe.

Let us first clarify some mathematical concepts and their relation to cosmic evolution. A point $\mathbf{x}_\mathrm{c}$ is said to be a fixed point or a critical point of the autonomous system if $f(\mathbf{x}_\mathrm{c})=0$. The fixed points are divided into three categories based on their stability. The first class is the stable node and stable spirals, distinguished by whether the eigenvalues of Jacobian matrix are pure real or complex. A stable fixed point $\mathbf{x}_\mathrm{c}$, no matter a stable node or a stable spiral, is an attractor. It represents the position where the system converges towards in the infinite future, i.e. $\mathbf{x}\to\mathbf{x}_\mathrm{c}$ for $t\to\infty$. In the context of cosmic evolution, the Universe eventually moves to an attractor in phase space. Point that indicates matter-dominated or radiation-dominated epoch should not be an attractor generally, since the Universe has already entered an accelerating stage. On the contrary, models with a stable fixed point representing the accelerating expansion are theoretically favored, especially those give the current radio of background fluid and dark energy for solving the coincidence problem. The second class refers to the unstable node, often considered as the initial point of evolution, and is thus sometimes termed the ``past attractor". The radiation dominated or matter dominated stage is encouraged to be unstable in order to initiate the observational acceleration. 
The last class is the saddle point, neither serving as a past attractor nor a future one. It may potentially represent a certain stage of cosmic evolution, but this stage is only temporary, as the system cannot actually reach the saddle point and therefore cannot remain there for an extended duration, unless the system initially starts there.

\begin{figure}[h]
\centering
\begin{minipage}{0.62\textwidth}
  \centering
  \includegraphics[width=\linewidth]{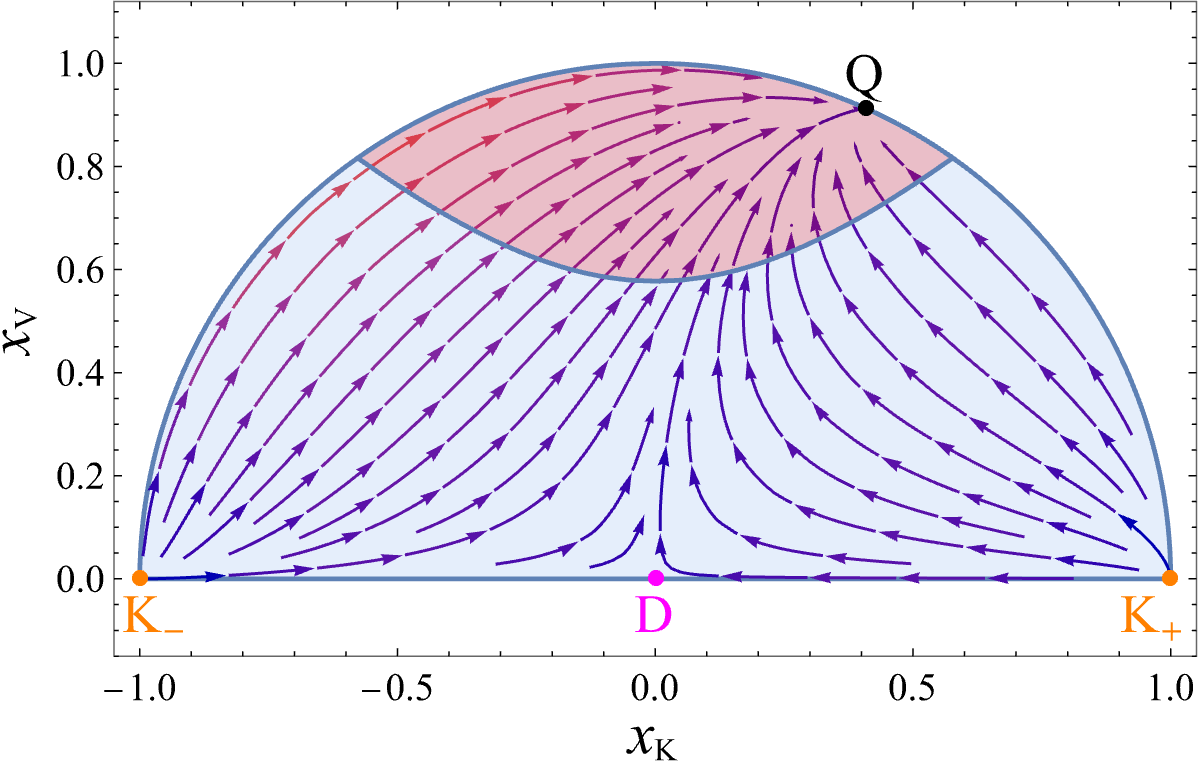}
\end{minipage}
\begin{minipage}{0.62\textwidth}
  \centering
  \includegraphics[width=\linewidth]{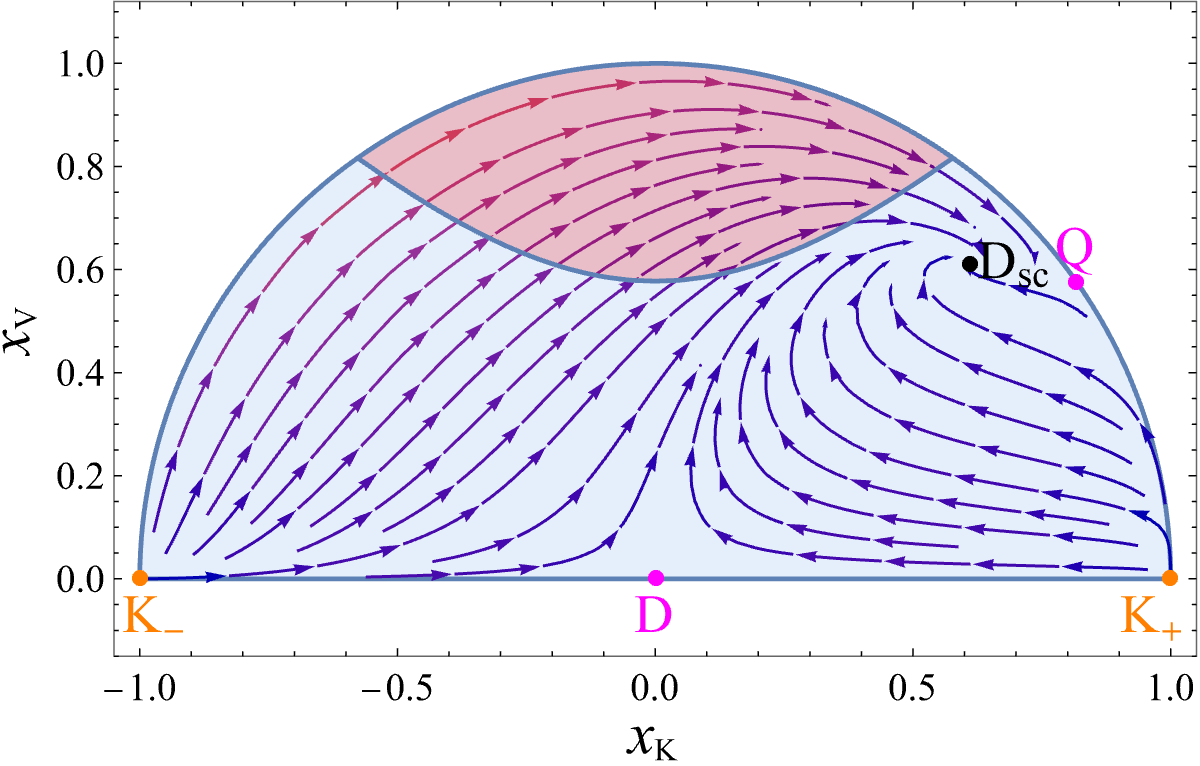}
\end{minipage}
\begin{minipage}{0.62\textwidth}
  \centering
  \includegraphics[width=\linewidth]{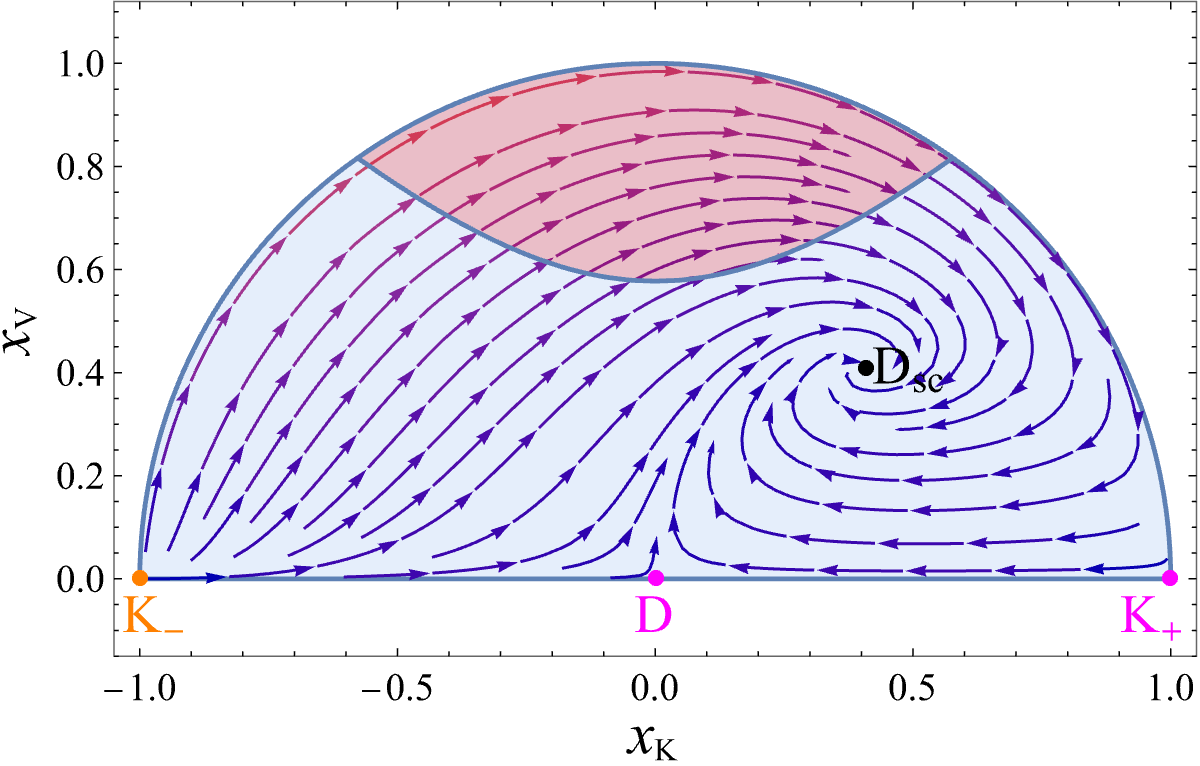}
\end{minipage}
\caption{The phase space of $x_\mathrm{K}$ and $x_\mathrm{V}$ with $\lambda=1$ (the upper subfigure), $\lambda=2$ (the middle subfigure), and $\lambda=3$ (the bottom subfigure). Here we assume $x_\mathrm{r}=0$ to reduce the phase space to two dimensions. The fixed points listed in \Tref{table1} are plotted, with yellow, magenta, and black representing stable, saddle, and unstable fixed points, respectively. Each streamline represents the evolution of the Universe under different initial conditions. The red region identifies accelerated expansion. It is clear that the scaling solution $\mathrm{D}_\mathrm{sc}$ is not in the accelerated expansion region.}
\label{fig:quint}
\end{figure}

By setting
$\rmd x_\mathrm{K}/\rmd N=\rmd x_\mathrm{V}/\rmd N=\rmd x_\mathrm{r}/\rmd N=0$ in Equations \eref{DS1}, \eref{DS2} and \eref{DS3}, the fixed points are acquired and those satisfying condition $$x_\mathrm{V}>0,\quad x_\mathrm{r}>0$$ are listed in \Tref{table1} and plotted in \Fref{fig:quint}.

\begin{table}
\caption{Properties of fixed points for the quintessence in the exponential
potential \eref{V} \cite{Copeland:1997et,Tsujikawa:2010sc}.}\label{table1}
\begin{indented}
\lineup
    \item[]\begin{tabular}{@{}lllllll}
    \br
    Fixed point	& $(x_\mathrm{K},x_\mathrm{V},x_\mathrm{r}$)&  $\Omega_\mathrm{r}$ & $\Omega_\mathrm{d}$ &   $\Omega_\mathrm{q}$ &$w_\mathrm{q}$
&   $w_\mathrm{eff}$\\
 \mr
R  &  $(0,0,1)$	&   $1$  &$0$ &   $0$ & --- &$\frac13$\\
D	& $(0,0,0)$&$0$&$1$&$0$&---&$0$\\
$\mathrm{R}_\mathrm{sc}$	&
$\Big(\frac{2\sqrt{6}}{3\lambda}, \frac{2\sqrt{3}}{3\lambda},
\frac{\sqrt{\lambda^2-4}}{\lambda}\Big)$ & $1-\frac{4}{\lambda^2}$ &
$0$&$\frac{4}{\lambda^2}$ & $\frac{1}{3}$ & $\frac13$\\
$\mathrm{D}_\mathrm{sc}$ &
$\Big(\frac{\sqrt{6}}{2\lambda},\frac{\sqrt{6}}{2\lambda},0\Big)$ & $0$
&$1-\frac{3}{\lambda^2}$& $\frac{3}{\lambda^2}$ & $0$ & $0$\\
$\mathrm{K}_\pm$  &  $(\pm1,0,0)$	&$0$& $0$ &  $1$   &   $1$ & $1$\\
Q   &  $\Big(\frac\lambda{\sqrt6},\frac{\sqrt{6-\lambda^2}}{\sqrt{6}},0$\Big)&
$0$& $0$ & $1$   &  $-1+\frac{\lambda^2}3$   &   $-1+\frac{\lambda^2}3$\\
\br
    \end{tabular}
\end{indented}
\end{table}

Point R (D) is a radiation (dust) dominated solution and is able to describe the
corresponding dominating epoch. Point $\mathrm{R}_\mathrm{sc}$
($\mathrm{D}_\mathrm{sc}$) is the scaling solution
\cite{Copeland:1997et,Ferreira:1997hj} for the ratio $\Omega_\mathrm{q}/\Omega_\mathrm{r}$
($\Omega_\mathrm{q}/\Omega_\mathrm{d}$) is a non-zero constant, which can be used to hide the presence of the scalar field in the cosmic evolution, at least at the background level. The scaling solution must satisfy
the condition $\lambda^2>3(1+w_\mathrm{m})$ in order to ensure that $\Omega_\mathrm{q}<1$. Since
$w_\mathrm{eff}=w_\mathrm{m}$ and usually $0<w_\mathrm{m}<1$, scaling solution usually does not
describe the cosmic acceleration (unless for a double exponential potential
\cite{Barreiro:1999zs} and some related others
\cite{Sahni:1999qe,Albrecht:1999rm,Dodelson:2000jtt}). Notice that the scaling solution returns to
the corresponding dominated solution in the limit $\lambda\to\infty$.

Points $\mathrm{K}_\pm$ represent solutions dominated by a scalar field but are
unable to account for the accelerating expansion due to the wrong EOS parameter.
Hence, the only fixed point that represents the present cosmic epoch is the
point Q, and the cosmic acceleration is realized if $\lambda^2<2$. As a result,
the evolution of the Universe is consistent with the trajectory $\mathrm R\to \mathrm
D\to \mathrm Q$.

\begin{figure}[h]
    \centering
    \includegraphics[width=0.75\textwidth]{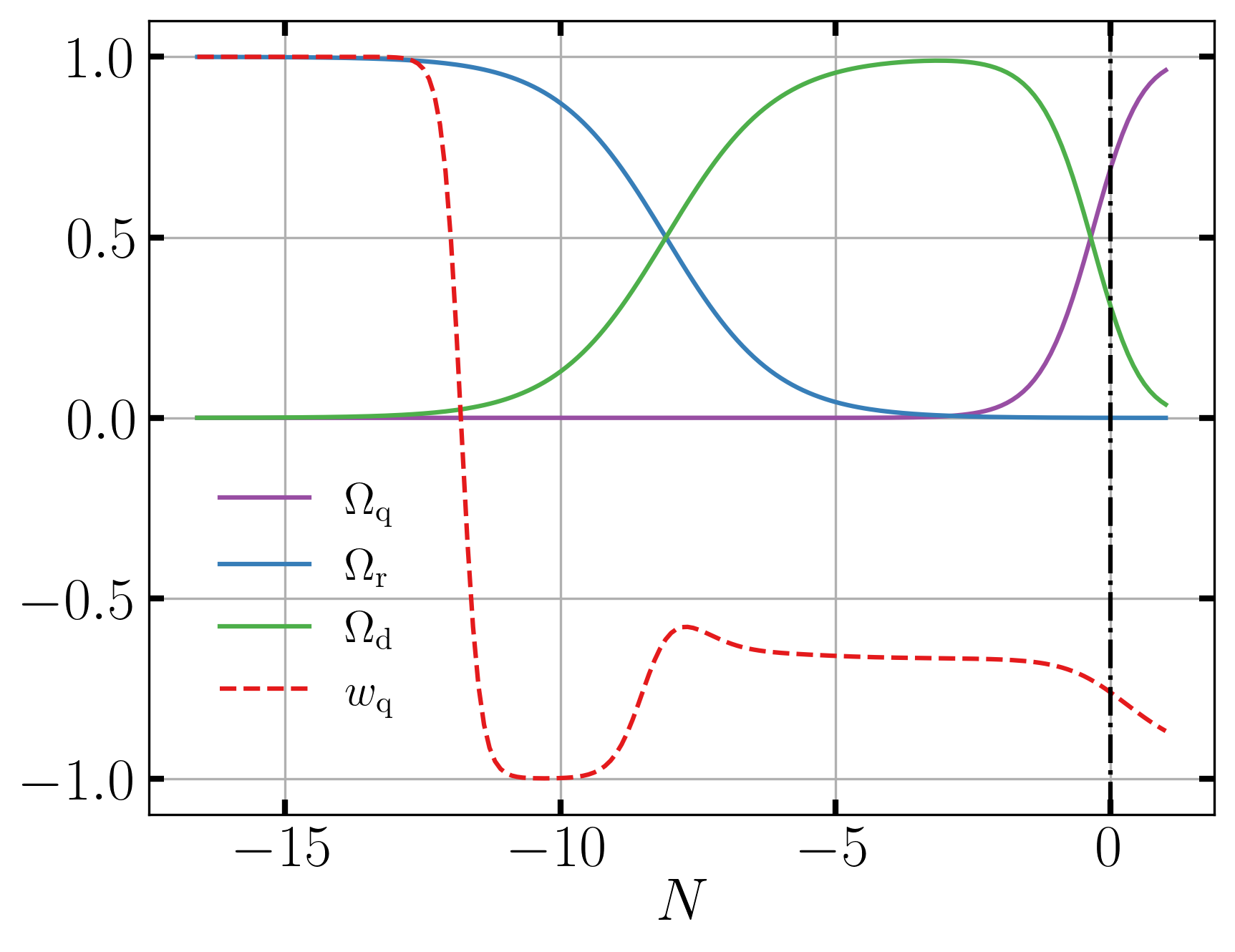}
    \caption{Evolution of $\Omega_\mathrm{r}$, $\Omega_\mathrm{d}$, $\Omega_\mathrm{q}$, and $w_\mathrm{q}$  for the potential $V(\phi)=M^5/\phi$ versus the $\rme$-folding time. The vertical dashed line denotes the present cosmological time, and the present values $\Omega_\mathrm{d}\simeq0.3$ and $\Omega_\mathrm{q}\simeq0.7$ are achieved. Here the complete curve of $w_\mathrm{q}$ is plotted, though it is literally no longer considered as quintessence when $w_\mathrm{q}>-1/3$. The solution enters the tracking regime in which the field energy density tracks the background fluid density. The initial conditions are chosen from \cite{amendola2010dark}.}
    \label{f:qO2}
\end{figure}

If $\lambda$ is not a constant, we have to consider an additional equation
\begin{equation}
	\frac{\rmd\lambda}{\rmd N}=-\sqrt{6}\lambda^2(\Gamma-1)x_\mathrm{K},
\end{equation}
where
\begin{equation}\label{gamma}
	\Gamma\equiv V\frac{\rmd^2V}{\rmd\phi^2}\left(\frac{\rmd V}{\rmd\phi}\right)^{-2}.
\end{equation}
In this scenario, the fixed points obtained in the case of a constant $\lambda$
can be considered as ``instantaneous'' fixed points that change over time
\cite{delaMacorra:1999ff,Ng:2001hs}, if assuming that the time scale for the
variation of $\lambda$ is significantly shorter than $H^{-1}$, the time
scale that produces significant changes in the scale of the Universe.

The possibility of a transition from the fixed point $\mathrm{D}_\mathrm{sc}$ to Q
arises when $\lambda$ decreases over time. If the condition $\Gamma>1$ is
satisfied, the absolute value of $\lambda$ decreases towards $0$. This means
that the solutions finally approach the accelerated instantaneous point Q even
if $\lambda^2>2$ during radiation and matter domination. The condition
$\Gamma>1$ is the so-called tracking condition under which the field density
eventually catches up with that of the background fluid, and a tracking solution
characterized by
\begin{equation}\label{TS}
	\Omega_\mathrm{q}\simeq\frac{3(1+w_\mathrm{q})}{\lambda^2}
\end{equation}
is called a tracker \cite{Steinhardt:1999nw}. \Fref{f:qO2} shows the evolution of each components of a tracking solution. If $\Gamma$ varies slowly in time,
the EOS parameter of quintessence is nearly constant during the matter
domination and radiation domination periods
\cite{Steinhardt:1999nw,Tsujikawa:2013fta},
\begin{equation}\label{wQ}
w_\mathrm{q}\simeq w_\mathrm{q}^\rmi=\frac{w_\mathrm{m}-2(\Gamma-1)}{2\Gamma-1}.
\end{equation}
The fact that $w_\mathrm{q}<w$ results in a slower decrease of the quintessence energy
density compared to the fluid energy density, and ultimately it leads to an
attainable value of $\Omega_\mathrm{q}=1$. From \Eref{TS}, $w_\mathrm{q}$ is
$-1+\lambda^2/3$ in the scalar field dominated epoch, which corresponds to the
stable fixed point Q as long as $\lambda<3(1+w_\mathrm{m})$. The tracker fields
correspond to attractor-like solutions, wherein the field energy density tracks
the background fluid density across a wide range of initial conditions \cite{Zlatev:1998tr}. This
implies that the energy density of quintessence does not necessarily need to be
significantly smaller than that of radiation or matter in the early Universe,
unlike in the cosmological constant scenario. Consequently, this offers a
potential solution to alleviate the coincidence problem. Another benefit of the
tracker solution is that it does not require the introduction of a new mass
hierarchy in fundamental parameters \cite{Zlatev:1998tr}.

In contrast to the tracking solution where $\Gamma>1$, if $\lambda$ gradually
increases over time, the accelerated phase of the Universe is limited at late
times because the energy density of the scalar field becomes negligible compared
to that of the background fluid.

\subsection{Cosmological Boundary}
\subsubsection{Phantom Dark Energy.}
The EOS parameter for the cosmological constant, $w_\Lambda=-1$, is referred to as the cosmological boundary, for the reason not only distinguishing between the rates of accelerating expansion, whether slower or faster than exponential, but also marking the point where perturbation divergence occurs \cite{Ma:1995ey}. A canonical scalar field, when treated as a perfect fluid, is bound by an EOS parameter within the range $(-1,1)$. However, current observations suggest the possibility of the equation of state parameter of dark energy being smaller than $w_\Lambda$, a scenario even favored by the data \cite{Corasaniti:2004sz,Alam:2003fg,Zhao:2006bt,Planck:2015fie}.

The phantom dark energy model was proposed as a complement to the canonical quintessence model \cite{Caldwell:1999ew}, offering a super-accelerating phase, the expansion faster then exponential. This can be achieved by substituting the quintessence action with an alternative one:
\begin{equation}\label{PA}
S_{\mathrm{p}}[g^{\mu\nu},\phi]=\int
\rmd^4x\sqrt{-g}\mathcal L_\mathrm{p}=\int
d^4x\sqrt{-g}\left[\frac12g^{\mu\nu}\partial_\mu\phi\partial_\nu\phi-V(\phi)\right].
\end{equation}
The EOS parameter of the phantom field is
\begin{equation}
w_\mathrm{p}\equiv\frac{p_\mathrm{p}}{\rho_\mathrm{p}}=\frac{\dot\phi^2+2V(\phi)}{\dot\phi^2-2V(\phi)},
\end{equation}
then one has $w_\mathrm{p}>1$ for the kinetic dominated regime ($\dot\phi^2/2>V$) and $w_\mathrm{p}<-1$ for the potential dominated regime ($V>\dot\phi^2/2$).

\begin{figure}[h]
    \centering
    \includegraphics[width=0.75\textwidth]{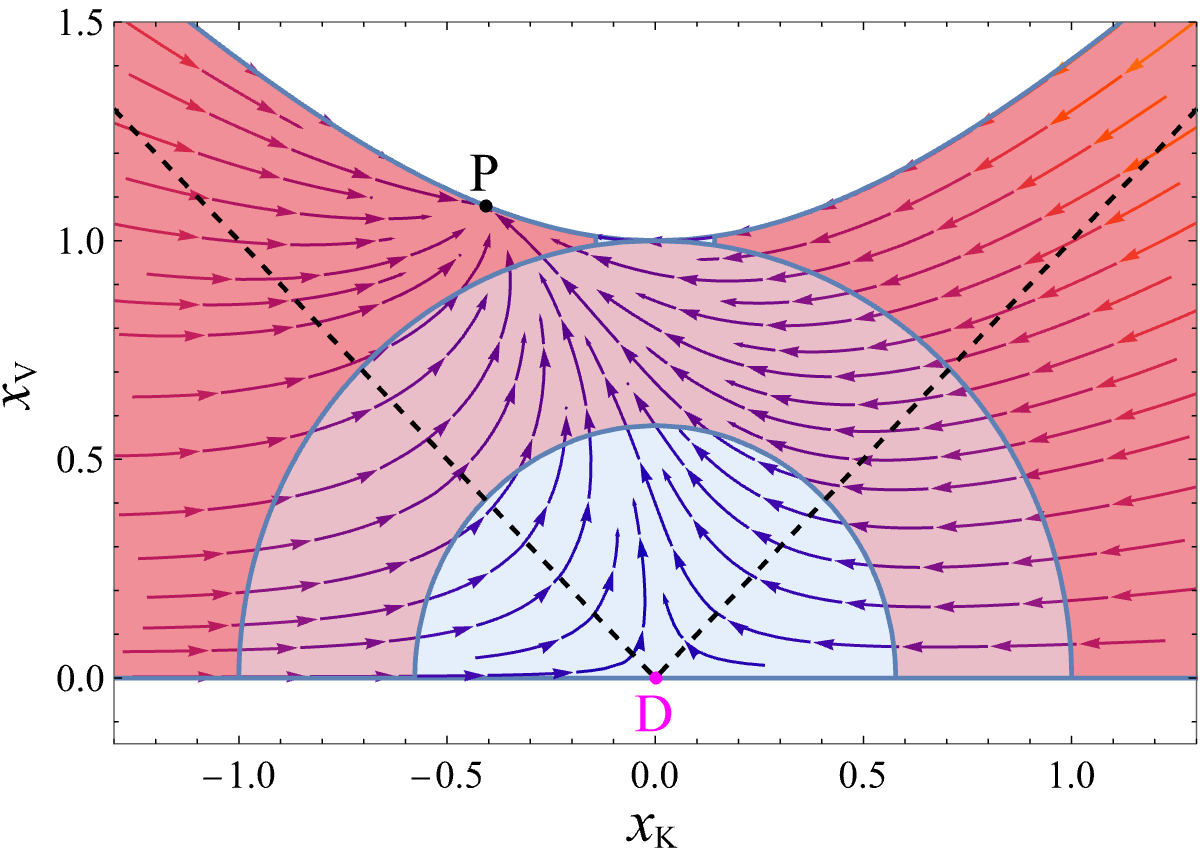}
    \caption{The phase space of $x_\mathrm{K}$ and $x_\mathrm{V}$ with value $x_\mathrm{r}=0$ and $\lambda=1$. The magenta saddle fixed point D represents the matter domination, while the black stable point P represents the phantom dominated stage. Each streamline represents the evolution of the Universe under different initial conditions. The light red region shows where Universe undergoes a standard accelerated expansion $-1<w_\mathrm{eff}<-1/3$, while the dark red region identifies expansion faster then exponential. The two green dashed lines correspond to $x_\mathrm{V}=\pm x_\mathrm{K}$ where the phantom EOS parameter diverges.}
    \label{fig:phdm}
\end{figure}

The choice of the same expansion normalised variables \eref{ds} also closes the system if the exponential potential \eref{V} is considered. Hence, the analogous analysis can be transplanted for phantom models. \Fref{fig:phdm} shows the phase space of a phantom dark energy Universe with $w_\mathrm{m}=0$ (a different $w_\mathrm{m}\in[0,\frac13]$ doesn't alter qualitative descriptions below). Only two fixed points exist in the system: one is the saddle point D, where the Universe contains only matter, and the other is the stable state P, representing a phantom dominated attractor. A heteroclinic orbit connecting point D to point P would represent the late time translation from matter to phantom domination. On the way to the attractor, the expanding solution for the scalar field will be
\begin{equation}
    a(t)\propto(t_\infty-t)^{-2/\lambda^2},
\end{equation}
ultimately leading to the occurrence of the ``Big Rip'' singularity within a finite proper time $t_\infty$ \cite{Caldwell:2003vq,Ellis_Maartens_MacCallum_2012}. A potential bound from above is necessary to suppress this singularity, since the phantom field rolls up the potential and subsequently converge towards $w_\mathrm{p}=-1$ as the field stabilizes at the maximum of potential \cite{Carroll:2003st,Singh:2003vx}.

Quantum instability always arises in phantom models due to the flipped sign of the kinetic term. Once the phantom quanta
interact with other fields, even though gravity, there will be an instability of
the vacuum because energy is no longer bound from below
\cite{Carroll:2003st,Cline:2003gs}. Moreover, phantom models are problematic at
the classical level because a strongly anisotropic CMB background is expected
unless the Lorentz symmetry is broken for the phantom at an energy scale lower
than 3 MeV \cite{Cline:2003gs}.
Last but not least, the phantom model with an exponential potential fails to address the coincidence problem and suffers from the fine-tuning of initial condition. Consequently, the inclusion of sophisticated potentials and matter fluids beyond the range of $[0,\frac13]$ have to be considered \cite{Urena-Lopez:2005pzi,Elizalde:2004mq}.

\subsubsection{Cross the Cosmological Boundary.}
The EOS parameters for dark energy in both the quintessence and phantom models are confined to opposite sides of the cosmological boundary, preventing them from crossing over into each other's domain. A no-go theorem indeed confirms this, stating as follows \cite{Xia:2007km}: for the theory of dark energy in the FLRW Universe described by a single fluid or a single scalar field $\phi$ with a Lagrangian $\mathcal{L}=\mathcal{L}(\phi,\partial_\mu\phi\partial^\mu\phi)$, which minimally coupled to Einstein gravity, its EOS $w_\mathrm{de}$ cannot across the cosmological constant boundary.

In a scenario where gravity is still governed by general relativity, additional degrees of freedom are necessary to enable dark energy to effectively traverse the boundary. The quintom model was proposed as a multi-field model incorporating a
quintessence-like field and a phantom-like field \cite{Feng:2004ad,Guo:2004fq},
\begin{equation}
    \fl S_\mathrm{qp}[g^{\mu\nu},\phi_\mathrm{q},\phi_\mathrm{p}]=\int\rmd^4x\sqrt{-g}\mathcal{L}_\mathrm{qp}\left[-\frac12g^{\mu\nu}\partial_\mu\phi_\mathrm{q}\partial_\nu\phi_\mathrm{q}+\frac12g^{\mu\nu}\partial_\mu\phi_\mathrm{p}\partial_\nu\phi_\mathrm{p}-V(\phi_\mathrm{q},\phi_\mathrm{p})\right].
\end{equation}
The energy density, pressure and the corresponding EOS parameter of the quintom model are
\begin{eqnarray}
    \rho_\mathrm{de}=\frac{1}{2}\dot\phi_\mathrm{q}^2-\frac12\dot\phi_\mathrm{p}^2+V(\phi_\mathrm{q},\phi_\mathrm{p}),\\
    p_\mathrm{de}=\frac{1}{2}\dot\phi_\mathrm{q}^2-\frac12\dot\phi_\mathrm{p}^2-V(\phi_\mathrm{q},\phi_\mathrm{p}),\\
    w_\mathrm{de}\equiv\frac{p_\mathrm{qp}}{\rho_\mathrm{qp}}=\frac{\dot\phi_\mathrm{q}^2-\dot\phi_\mathrm{p}^2-2V(\phi_\mathrm{q},\phi_\mathrm{p})}{\dot\phi_\mathrm{q}^2-\dot\phi_\mathrm{p}^2+2V(\phi_\mathrm{q},\phi_\mathrm{p})}.
\end{eqnarray}
It is clear that $w_\mathrm{qp}\gtrless-1$ if $\dot\phi_\mathrm{q}\gtrless\dot\phi_\mathrm{p}$ and the cosmological constant scenario is recovered when $\dot\phi_\mathrm{q}=\dot\phi_\mathrm{p}$.

The potential $V(\phi_\mathrm{q},\phi_\mathrm{p})$ for non-interacting scalar fields can be generally written as
\begin{equation}
    V(\phi_\mathrm{q},\phi_\mathrm{p})=V_\mathrm{q}(\phi_\mathrm{q})+V_\mathrm{p}(\phi_\mathrm{p}),
\end{equation}
where $V_\mathrm{q}(\phi_\mathrm{q})$ and $V_\mathrm{p}(\phi_\mathrm{p})$ are arbitrary self-interacting potentials for the quintessence and phantom respectively. In such case, the two scalar field can be treated exactly as if they were single models. Due to the fact that $w_\mathrm{p}<w_\mathrm{q}$, the Universe will eventually evolve to a stage where phantom dominates $w_\mathrm{eff}=w_\mathrm{de}\le-1$. The $w_\mathrm{de}$ across the cosmological boundary from above to below if the matter dominated epoch is followed by a temporary quintessence dominated stage \cite{Guo:2004fq,Leon:2012vt,Cai:2009zp}. If future observations will constrain the EOS parameter of dark energy to be below $-1$ today but above $-1$ in the past, then the quintom scenario of is arguably the simplest framework where such a situation can arise.

Unfortunately, neither the coupled nor the uncoupled quintom models seem to solve the cosmic coincidence problem and the fine-tuning of initial conditions simultaneously, not to mention the fundamental problems associated with the ghost field. Furthermore, couplings between kinetic terms and scalar fields, like $\partial_\mu\phi_\mathrm{q}\partial^\mu\phi_\mathrm{p}$ \cite{Saridakis:2009jq} and the so-called hessence dark energy model \cite{Wei:2005nw,Wei:2005fq,Alimohammadi:2006qi} are investigated as well. Finally, theories beyond general relativity, such as those incorporating higher curvature terms and non-minimally coupled scalar fields between gravity or matter, can also cross the cosmological boundary without encountering negative kinetic energy and the associated quantum instability, as discussed in \ref{412}.

\subsection{Non-canonical Kinetic Term}
In the preceding sections, some simplest dynamical dark energy models are introduced, providing as much detail as possible regarding their evolution. Here we briefly review dark energy models from scalar fields with modified kinetic terms, which hold significant potential for addressing those problems of $\Lambda$CDM model at once.

\subsubsection{k-essence.}
A scalar field model of dark energy with a modified kinetic term is the
so-called kinetically driven essence, or k-essence for simplicity.
The initial focus of the k-essence
was on its application to inflation \cite{Armendariz-Picon:1999hyi}, and numerous theoretical models, such as low energy effective string theory \cite{Gasperini:2002bn}, ghost condensate \cite{Arkani-Hamed:2003pdi}, tachyon \cite{Garousi:2000tr,Sen:2002nu}, and Dirac-Born-Infeld (DBI) theory, \cite{Silverstein:2003hf,Alishahiha:2004eh} can be classified under this
dark energy model.

The action of the k-essence dark energy model is given by \cite{Armendariz-Picon:1999hyi}
\begin{equation}\label{KEA}
\fl S[g^{\mu\nu},\phi,\Psi_\mathrm{m}]=S_\mathrm{EH}[g^{\mu\nu}]+\int
\rmd^4x\sqrt{-g}P(\phi,X)+S_\mathrm{m}[g^{\mu\nu},\Psi_\mathrm{m}],
\end{equation}
where $P(\phi,X)$ is an arbitrary function of a scalar field $\phi$ and its
kinetic energy 
\begin{equation}
	X\equiv-\frac12g^{\mu\nu}\partial_\mu\phi\partial_\nu\phi.
\end{equation}
The k-essence model can achieve accelerated expansion solely through kinetic energy, even in the absence of a potential \cite{Chiba:1999ka,Malquarti:2003nn}, distinguishing it from models with canonical kinetic energy.
 
The energy-momentum tensor of the k-essence is given by
\begin{equation}
T^\mathrm{k}_{\mu\nu}=-\frac{2}{\sqrt{-g}}\frac{\delta(\sqrt{-g}P)}{\delta g^{\mu\nu}} =
P_X\partial_\mu\phi\partial_\nu\phi+g_{\mu\nu}P,
\end{equation}
where the subscript $X$ of function $P$
represents the partial derivative with respect to $X$.
Comparing with the energy-momentum tensor of a perfect fluid \eref{Tpf}, we
have the pressure and energy density of the k-essence,
\begin{eqnarray}
	p_\mathrm{k}=P,\\
	\rho_\mathrm{k}=2XP_X-P,
\end{eqnarray}
with the 4-velocity $u_\mu=\partial_\mu\phi/\sqrt{2X}$. If $P$ only depends on
$\phi$, we obtain $p_\mathrm{k}=-\rho_\mathrm{k}$, which corresponds to general relativity with the cosmological constant.
If $P$ solely depends on $X$, we have $\rho_\mathrm{k}=\rho_\mathrm{k}(X)$. Hence, $\rho_\mathrm{k}$ can be
rearranged to give $p_\mathrm{k}=p_\mathrm{k}(\rho_\mathrm{k})$, which is the EOS for an isentropic fluid.
In the general case where $p_\mathrm{k}=P(\phi,X)$, the hydrodynamical analogy is still
useful, and the EOS parameter is acquired directly,
\begin{equation}\label{WK}
w_\mathrm{k}\equiv\frac{p_\mathrm{k}}{\rho_\mathrm{k}}=\frac{P}{2XP_X-P},
\end{equation}
and $w_\mathrm{k}$ approaches to $-1$ if the condition $|XP_X|\ll|P|$ is satisfied.

A non-canonical scalar field Lagrangian will in general introduce theoretical problems at both the quantum and classical levels. Hence, the selection of function $P$ has limitations. The propagation sound speed $c_\mathrm{s}^\mathrm{k}$ of the k-essence field is given by
\cite{Garriga:1999vw},
\begin{equation}\label{KCS}
\big(c^{\mathrm{k}}_\mathrm{s}\big)^2=\frac{\partial p_k/\partial X}{\partial \rho_k/\partial
X}=\frac{P_X}{P_X+2XP_{XX}},
\end{equation}
which should be a positive number.
Furthermore, due to the positive definiteness of the Hamiltonian \cite{Piazza:2004df}, both the denominator $\xi_1$ and the numerator $\xi_2$ of $\big(c^{\mathrm{k}}_\mathrm{s}\big)^2$ in Equation (\ref{KCS}) should be positive.
To be in accordance with the principle of causality, $c_\mathrm{s}^\mathrm{k}$ is considered favorable if
it does not exceed the speed of light which gives $P_{XX}\geq0$; however, see
Ref.~\cite{Babichev:2007dw}.

On account of the unknown dependence upon the arbitrary function $P(\phi,X)$, it is too complicated to directly study with dynamical systems techniques. 
Therefore, to study their dynamics and evolution, k-essence are typically divided into several classes \cite{Bahamonde:2017ize}:
\begin{eqnarray}
    \mathrm{Class\ I:}\quad &P(\phi,X)=XG(Y),\quad Y=\frac{X}{V(\phi)},\label{kI}\\
    \mathrm{Class\ II:}\quad &P(\phi,X)=K(\phi)\tilde{P}(X),\label{kII}\\
    \mathrm{Class\ III:}\quad &P(\phi,X)=F(X)-V(\phi),\label{kIII}
\end{eqnarray}
where $G$, $K$, $\tilde P$, $F$ and $V$ are functions of their own arguments. It is worth noting that a model can be classified into one of two classes, as seen with the quintessence model, or it may not fit into either category, as exemplified by the DBI model.

The first class of k-essence exhibits a deep connection with scaling solutions that are of great relevance for the cosmic coincidence problem. It has been proved that \cite{Piazza:2004df} scaling solutions appear in k-essence models only if a Lagrangian of the type \eref{kI} is assumed with $V(\phi)\propto\rme^{-\lambda\phi/M_\mathrm{Pl}}$. Further research indicated that dark energy late-time solutions are invariably unstable in the presence of a scaling solution, unless they belong to the phantom type, even with a coupling to the matter sector \cite{Tsujikawa:2006mw,Amendola:2006qi}. Note that the quintessence exhibits scaling solutions because it corresponds to $G(Y)=1-1/Y$.

Unlike the Class I, the second class is intriguing for its tracking solutions. It has been shown that tracking solutions naturally appear for the models \eref{kII} during the radiation dominated era, and a cosmological-constant-like behavior shortly after the transition to matter domination were put forward, which claims to have solved
the coincidence problem without fine-tuning the parameters \cite{Armendariz-Picon:2000nqq,Armendariz-Picon:2000ulo}. However, a singularity associated with a diverging sound speed is present in such models \cite{Bonvin:2006vc}, so that it cannot arise as a low-energy effective field theory of a causal, consistent, high-energy theory. Moreover, the Lagrangians in these models are hard to construct in the framework of particle physics. 

The third class \eref{kIII} has a non-canonical kinetic term appears together with a standard self-interacting potential for the scalar field. The famous (dilatonic) ghost condensate is usually considered a subclass of this kind, which is simply recommended in next \Sref{dGC}. One interesting thing is that some special functions $F(X)$ and $V(\phi)$ produce scaling solutions, even if the Lagrangian cannot be written in the form of Class I \cite{De-Santiago:2012ibi}.

\subsubsection{(Dilatonic) Ghost Condensate.}\label{dGC}
The quintessence and phantom models, as special cases of k-essence, have
\begin{equation}
    P(\phi,X)=\epsilon X-V(\phi),
\end{equation}
where $\epsilon=\pm1$ represents quintessence and phantom respectively.

The denominator and numerator of propagation sound speed are $\xi_1=\xi_2=\epsilon$, indicating the phantom model is unstable.  The instabilities come form the perturbation level, where the gradient energy plays an important role. Therefore, one approach is to introduce a background action resembling that of a ghost, allowing the spatial gradient terms to address the instability issue. To implement the idea, Arkani-Hamed et al. \cite{Arkani-Hamed:2003pdi} proposed an effective field theory of a rolling ghost, which can be conceptualized as an infrared modification of gravity. They termed this modification as ghost condensate with the arbitrary function taking the form
\begin{equation}
    P(\phi,X)=-X+\frac{X^2}{M^4},
\end{equation}
where $M$ is a constant having a dimension of mass. This model was later generalized to a modified version as
\begin{equation}
    P(\phi,X)=-X+\frac{X^2}{M^4}\exp{\left(\frac{\lambda\phi}{M_\mathrm{Pl}}\right)},
\end{equation}
which is called dilatonic ghost condensate model \cite{Piazza:2004df}.

Here we briefly introduce the cosmic evolution of dilatonic ghost condensate dark energy model. The pressure, energy density and the corresponding EOS parameter are
\begin{eqnarray}
    p_\mathrm{dGC}=-X+\frac{X^2}{M^4}\rme^{\lambda\phi},\\
    \rho_\mathrm{dGC}=-X+\frac{3X^2}{M^4}\rme^{\lambda\phi},\\
    w_\mathrm{dGC}\equiv\frac{p_\mathrm{dGC}}{\rho_\mathrm{dGC}}=\frac{1-\rme^{\lambda\phi}X/M^4}{1-3\rme^{\lambda\phi}X/M^4},
\end{eqnarray}
where $M_\mathrm{Pl}=1$ is used for briefness. Quantum stability necessitates $\rme^{\lambda\phi}X/M^4 \geq 1/2$, resulting in the EOS parameter lying within the range $-1 \leq w_\mathrm{dGC} <1/3$. In particular the de Sitter solution $w_\mathrm{dGC}=-1$ is realized at $X=M^4e^{-\lambda\phi}/2$. Then it is possible to explain the present cosmic acceleration for $M\sim10^{-3}\;\mathrm{eV}$ in ghost condensate model as $\rho_\mathrm{GC}=-p_\mathrm{dGC}=M^4/4$ at the de Sitter point. Hence, the (dilatonic) ghost condensate is another example of accelerated expansion that can be generated without the assistance of a potential. 

By adjusting the selection of dynamic variables to \cite{amendola2010dark}
\begin{equation}
    x_\mathrm{K}\equiv\frac{\dot\phi}{\sqrt{6}H},\quad x_\mathrm{M}\equiv\frac{\dot\phi^2\rme^{\lambda\phi}}{2M^4},\quad x_\mathrm{r}\equiv\frac{\sqrt{\rho_\mathrm{r}}}{\sqrt{3}H}.
\end{equation}
we can get the autonomous system describing the dilatonic ghost condensate model. There always exists a stable point describing accelerated expansion toward which the Universe evolves, as long as $0 \leq \lambda < \sqrt{6}/3$. The corresponding sound speed is in the range $0\le c_\mathrm{s}^\mathrm{dGC}<1/3$, which means this model does not violate causality. The EOS parameter of this future attractor is \cite{amendola2010dark}
\begin{equation}
    w_\mathrm{eff}=w_\mathrm{dGC}=\frac{\lambda^2f(\lambda)-8}{3\lambda^2f(\lambda)+8},\quad f(\lambda)=1+\sqrt{1+\frac{16}{3\lambda^2}},
\end{equation}
which suggests that the ghost condensate behaves like a cosmological constant in dynamics during the final stage of cosmic evolution.

\subsubsection{Unified Dark Matter and Dark Energy.}
The temptation to unify dark matter and dark energy into a single entity has occurred to many cosmologists almost from the beginning, in order to solve two big dark sector problems at once. Both a fluid and a scalar field are attempted to represent dark matter and dark energy at the same times \cite{Bento:2002ps,Scherrer:2004au,Bertacca:2007ux,Fukuyama:2007sx}. Here we simply introduce a kind of k-essence model that is regarded as dark matter and dark energy simultaneously.

The Lagrangian of a such k-essence model with only a kinetic term is \cite{Scherrer:2004au}
\begin{equation}
    P(\phi,X)=-F_0+F_2(X-X_0)^2,
\end{equation}
where $F_0$ and $F_2$ are constants. Notice that there is an extremum $X_0$ of the kinetic energy, around which the pressure and energy density of k-essence are approximately
\begin{eqnarray}
    p_\mathrm{k}\simeq-F_0,\\
    \rho_\mathrm{k}\simeq F_0+4F_2X_0(X-X_0).
\end{eqnarray}
Furthermore, by substituting $X=X_0(1+\varepsilon a^{-3})$ with $0<\varepsilon a^{-3}\ll1$ ($\varepsilon$ being a constant), the solution to the continuity equation around $X_0$ yields the EOS parameter
\begin{equation}
    w_\mathrm{k}=-\left(1+\frac{4F_2}{F_0}X_0^2\varepsilon a^{-3}\right)^{-1}.
\end{equation}

It is possible to realize $w_\mathrm{k}\simeq0$ during the matter domination provided that the condition $4F_2X_0^2/F_0\gg1$ is satisfied, while $w_\mathrm{k}$ approaches the de Sitter value $-1$ at late time due to $\varepsilon a^{-3}\to0$. This specific model does not suffer from the problem of stability or causality if $X>X_0$, rendering it a more reliable unified model.

\subsection{Interacting Dark Energy Model}
Typically, the dark energy models are based on scalar fields minimally coupled to the gravity, and do not implement the explicit coupling of the field to background matter, as illustrated by the models introduced above. However, there is no fundamental reason for this assumption in the absence of an underlying symmetry supposed to suppress the coupling, in particular the dark sector that we barely no nothing about from a microscopic and dynamical field theory perspective. Although the interaction between dark energy and normal matter particles are heavily restricted by observations \cite{ParticleDataGroup:2014cgo,Peebles:2002gy}, this is not the case for dark matter particles, and the neglect of this potential interaction of dark components may result in misinterpretations of observational data. In fact, the study of the interaction between dark energy and dark matter is getting more promising recently, for the possible solution to the coincident problem and tensions in $\Lambda$CDM models.

\subsubsection{Phenomenological Description.}
Since there is no established approach from first principles that can specify the form of coupling between dark energy and dark matter, dark matter could manifest as either bosonic or fermionic particles, adhering to either the standard model or extending beyond it. Likewise, dark energy could be conceptualized as a fluid, a field, or in other forms. Constructing phenomenological models initially, grounded in intuition and experience, is consistently beneficial. These models can then be rigorously tested against diverse observational data.

The interaction between dark matter and dark energy is described by following modified conservation equations,
\begin{eqnarray}
    \dot\rho_\mathrm{dm}+3H\rho_\mathrm{dm}=Q,\label{IDE1}\\
    \dot\rho_\mathrm{de}+3H(1+w_\mathrm{de})\rho_\mathrm{de}=-Q,\label{IDE2}
\end{eqnarray}
where $Q$ is termed the interacting kernel and its sign determines the direction of the energy flux. Specifically, energy transfers from dark energy to dark matter when $Q>0$ and the transmission reverses when $Q<0$. The original model is restored if the interaction is ``turned off".

In order to illustrate how interaction between the dark components acts on cosmological dynamics, consider the time evolution of the radio $r\equiv\rho_\mathrm{dm}/\rho_\mathrm{de}$,
\begin{equation}
    \dot r=\frac{\rho_\mathrm{dm}}{\rho_\mathrm{de}}\left(\frac{\dot\rho_\mathrm{dm}}{\rho_\mathrm{dm}}-\frac{\dot\rho_\mathrm{de}}{\rho_\mathrm{de}}\right)=-3\Gamma Hr,\quad\Gamma=-w_\mathrm{de}-\frac{1+r}{\rho_\mathrm{dm}}\frac{Q}{3H}.
\end{equation}

It is not difficult to observe that the interaction kernel should be the product of an energy density and a term with an inverse time dimension. Additionally, it is natural to assume that the kernel depends solely on relevant dynamical quantities (such as $\rho_\mathrm{dm}$, $\rho_\mathrm{de}$ and $a$) and possibly their derivatives. The simplest cases can be the following kernels
$$Q\supset\xi_1\rho_\mathrm{dm}H,\xi_2\rho_\mathrm{de}H,\xi_3\dot\rho_\mathrm{dm},\xi_4\dot\rho_\mathrm{de},\cdots$$
and their linear combinations. Other forms of interaction kernel including nonlinearity and a lot of complexity are summarized in review papers like \cite{Wang:2016lxa,Wang:2024vmw}.

\begin{figure}[h]
    \centering
    \includegraphics[width=0.75\textwidth]{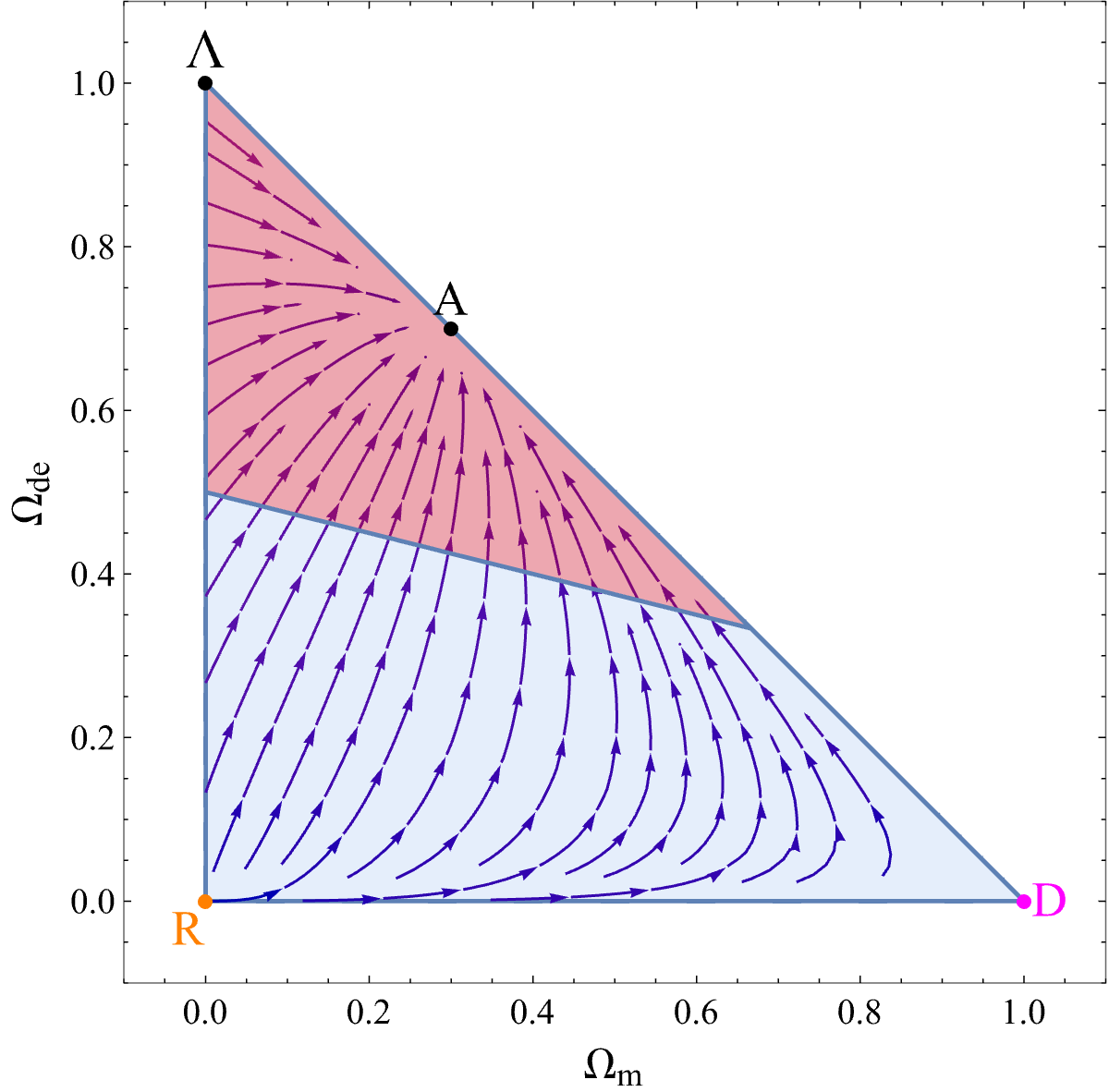}
    \caption{The diagram of $\Omega_\mathrm{m}$-$\Omega_\mathrm{de}$ with values $w_\mathrm{de}=-1$ and $\xi_2=0.3$ in the case where $Q=3\xi_2H\rho_\mathrm{de}$. The black point A is the only stable fixed point in the system representing the future attractor with $\Omega_\mathrm{m}=0.3$ and $\Omega_\mathrm{de}=0.7$. The black point $\Lambda$ is the future attractor of $\Lambda$CDM model with $\Omega_\mathrm{m}=0$ and $\Omega_\mathrm{de}=1$. The orange point R and magenta point D are unstable radiation domination and saddle matter domination respectively. Each streamline represents the evolution of the Universe under different initial conditions. The red region identifies accelerated expansion.}
    \label{f:IDE}
\end{figure}

Now, let us consider the case $Q=3H(\xi_1\rho_\mathrm{dm}+\xi_2\rho_\mathrm{de})$ and examine how the interacting dark energy model can resolve the coincidence problem. For this particular kernel,
\begin{equation}
    \Gamma=-w_\mathrm{de}-\frac{(1+r)(\xi_1r+\xi_2)}{r},
\end{equation}
and the stationary solutions are obtained imposing $r_\mathrm{s}\Gamma(r_\mathrm{s})=0$, which gives ($\xi_1\neq0$)
\begin{equation}
    r_\mathrm{s}^\pm=-\frac{w_\mathrm{de}+\xi_1+\xi_2\pm\sqrt{(w_\mathrm{de}+\xi_1+\xi_2)^2-4\xi_1\xi_2}}{2\xi_1}.
\end{equation}
The coincidence problem is completely solved if one of the above roots $r_\mathrm{s}^\pm$ is the future attractor of the system. Moreover, the evolution of the Universe should gradually approach the attractor from above, implying that the ratio of matter to dark energy decreases over time. In the special case $\xi_1=\xi_2=\xi$, the larger root $r_\mathrm{s}^-$ is a past attractor while the smaller root $r_\mathrm{s}^+$ is a future attractor \cite{Zimdahl:2001ar,Chimento:2003iea}. As Universe expands, $r(t)$ will evolve from $r_\mathrm{s}^-$ to the stable solution $r_\mathrm{s}^+$ avoiding the coincidence problem if $r_\mathrm{s}^+\sim 3/7$. If $\xi_2=0$, the system still retains a past attractor $r_s^+$ with no future attractor ($r_\mathrm{s}^-=0$ in this case). However, the variation is $|(\dot r/r)_0|<H_0$, slower than in $\Lambda$CDM, which mitigates the coincidence problem \cite{delCampo:2006vv}. As for the $\xi_1=0$ case, it is fortunate that the only fixed point $r_\mathrm{s}=-\xi_2/(w_\mathrm{de}+\xi_2)$ is stable. Consequently, there is no more coincidence problem if we select $\xi_2\sim0.3$ under the assumption of $w_\mathrm{de}=-1$. \Fref{f:IDE} showing the phase space of this example will help understand the conclusion.

While the initial drive behind interacting dark energy models was to solve or alleviate the coincidence problem, their focus has recently shifted towards elucidating the disparity between the observed value of the Hubble constant derived from CMB data and the local measurements. The interacting dark energy model increases the proportion of late-time dark energy, directly boosting the Hubble expansion rate. Moreover, the reduction in dark matter entails an increase in the Hubble constant to uphold the physical ratio of dark matter energy density in accordance with CMB constraints. In addition, dark matter decaying into dark energy also reduces structure growth of matter at late time, thereby alleviating the $S_8$ problem. Consequently, the interacting dark energy model emerges as a multifaceted potential solution to the Hubble tension.

\subsubsection{Coupled Scalar Fields.}
If focusing our attention on dark energy in the form of quintessence, one has the coupled quintessence model with the modified Klein-Gordon equation
\begin{equation}
    \ddot\phi+3H\dot\phi+\frac{\rmd V}{\rmd\phi}=-\frac{Q}{\dot\phi}.
\end{equation}
The basic kernel $Q\propto\rho_\mathrm{dm}\dot\phi$ naturally emerges in certain modified gravity theories (such as $f(R)$ gravity \cite{DeFelice:2010aj,He:2011qn}) in the Einstein frame, through which dynamics are introduced in \Sref{s421}.

In general, all aforementioned dark energy models can be generalized by introducing a coupling in the dark sector, following the same approach as with quintessence. The coupled phantom models with exponential potentials and frequently analyzed couplings demonstrate that scaling solutions cannot remain stable \cite{Copeland:2006wr,Shahalam:2017fqt}, as is the case with quintessence. Therefore, they are unable to resolve the cosmic coincidence problem.

The situation is different for more general k-essence models. Scaling solutions typically exist in Lagrangians of the form $P(\phi,X)=Q^2(\phi)XG(XQ^2(\phi)\rme^{\lambda\phi/M_\mathrm{Pl}})$, where $Q(\phi)$ is the coupling function and $G$ an arbitrary function \cite{Amendola:2006qi}. However, it has been demonstrated that an evolutionary trajectory capable of resolving the coincidence problem is prohibited due to the singularity associated with both dynamical variables and the sound speed \cite{Amendola:2006qi}. As additional resources, the study of coupled tachyonic models can be found in \cite{Gumjudpai:2005ry,Farajollahi:2011jr,Farajollahi:2011ym}, while coupled DBI scalar field models are discussed in \cite{Mahata:2015lja,Kaeonikhom:2012xr}.

\section{Scalar-Tensor Gravity}\label{Sec4}

Physicists introduced dark energy to explain the accelerating phase of the
Universe, but the nature of this bizarre substance is unknown and the validity
has also been called into questions. Since the evidence for dark energy comes
entirely from its gravitational effects, which are inferred assuming the
validity of general relativity, the consideration of modifications to general
relativity, such that the necessity of dark energy is obviated, is another rational
approach~\cite{Clifton:2011jh}. In this section, we  focus on the scalar-tensor
gravity, one of the most studied modified theories of gravity.

\subsection{Introduction to Scalar-Tensor Gravity}

The central idea behind scalar-tensor gravity theory is the incorporation of a
scalar field alongside the metric tensor field to participate in the
gravitational interaction. This addition of the scalar field allows for
variations in the strength of gravity over spacetime, opening up new avenues for
understanding the fundamental forces of the Universe. Scalar-tensor theory has
since evolved and been refined in various forms and models, becoming a focal
point for research in cosmology, astrophysics, and fundamental
physics~\cite{Fujii:2003pa}.

The origins of scalar-tensor gravity can be traced back to the pioneering work
in the mid-20th century. One of the earliest proponents of this theory was the
German physicist Jordan \cite{jordan1955schwerkraft}, who in 1955 introduced the concept
of a scalar field coupled to gravity as a means to unify gravity with
electromagnetism. Moreover, it was Brans and Dicke who further developed and
formalized the theory in 1961 \cite{Brans:1961sx}, now known as the Brans-Dicke
(BD) theory. The action for the theory is,
\begin{equation}\label{BDA}
	\fl S_\mathrm{BD}[g^{\mu\nu},\Phi,\Psi_\mathrm{m}]=\int \rmd^4x\sqrt{-g}\left(\Phi
	\frac{R}{2\kappa}-\frac{\omega_\mathrm{BD}}{2\Phi} g^{\mu\nu}
	\partial_\mu\Phi\partial_\nu\Phi\right)+S_\mathrm{m}[g^{\mu\nu},\Psi_\mathrm{m}],
\end{equation}
where $w_\mathrm{BD}$ is the only parameter in the theory. Here we add a
coefficient of $\frac12$ to the Lagrangian of the gravitational part of the
traditional BD theory. Shortly afterward, Bergmann \cite{Bergmann:1968ve} and
Wagoner \cite{Wagoner:1970vr} separately generalized the action to
\begin{equation}\label{JFA}
	\fl S_\mathrm{J}[g^{\mu\nu},\Phi,\Psi_\mathrm{m}] = \int \rmd^4x\sqrt{-g}\left[\Phi
	\frac
	R{2\kappa}-\frac{W(\Phi)}{2\Phi}g^{\mu\nu} \partial_\mu
	\Phi\partial_\nu\Phi-U(\Phi)\right]+S_\mathrm{m}[g^{\mu\nu},\Psi_\mathrm{m}] \,,
\end{equation}
by letting the coupling parameter $w_\mathrm{BD}$ be a function of the scalar
field $W(\Phi)$ and adding a potential term $U(\Phi)$. The subscript ``J''
here represents the Jordan frame.

A suitable conformal transformation
\begin{equation}\label{CT}
	g_{\mu\nu}\to g_{\mu\nu}^*=A^{-2}(\phi)g_{\mu\nu}
\end{equation}
is able to take us from the Jordan frame to the Einstein
frame~\cite{Damour:2007uf}. We denote the scalar field in the Einstein frame as
$\phi$ instead of the field $\Phi$ in the Jordan frame. The action in the
Einstein frame is
\begin{equation}\label{EFA}
	\fl S_\mathrm{E}[g^{\mu\nu}_*,\phi,\Psi_\mathrm{m}]=\int
	\rmd^4x\sqrt{-g_*}\left[\frac{R_*}{2\kappa_*}-\frac12g_*^{\mu\nu}
	\partial_\mu\phi\partial_\nu\phi-V(\phi)\right]+S_\mathrm{m}[A^{-2}(\phi)
	g^{\mu\nu}_*,\Psi_\mathrm{m}],
\end{equation}
where $\kappa_*=8\pi G_*$, and $G_*$ is a bare gravitational constant. The
physical quantity marked with a star indicates that it is in the Einstein frame.
The potential $V(\phi)$ in the Einstein frame is related to the potential
$U(\Phi)$ in the Jordan frame by the rescaling function $A(\phi)$ as
\begin{equation}
	V(\phi)=A^4(\phi)U(\phi).
\end{equation}

The gravity theory is usually not invariant under conformal transformation,
including Einstein's general relativity. A natural question arises: which
conformal frame is physical? Unfortunately, as of now, the answer remains
unknown. In the subsequent \Sref{subsec1}
and \Sref{subsec2}, the Einstein frame and Jordan frame are both considered,
respectively. Cosmologies within these two frames are discussed as well in corresponding sections.

\subsection{Einstein Frame: Coupled Quintessence}\label{subsec1}

The canonical form of action \eref{EFA} is acquired if $A(\phi)$ satisfies
\begin{eqnarray}
	A^2(\phi)=\frac{\kappa}{\kappa_*\Phi},\label{CT1}\\
	\alpha^2(\phi)=\frac1{2[2\kappa W(\Phi)+3]},\label{CT2}
\end{eqnarray}
where $\alpha(\phi)\equiv{\rmd\ln A}/{\rmd\phi}$. The relation between scalar fields
$\Phi$ and $\phi$ in Jordan and Einstein frames is easily acquired from
Equations \eref{CT1} and \eref{CT2}, as
\begin{equation}
	\frac{\rmd\phi}{\rmd\Phi}=\pm\frac{\sqrt{\kappa W(\Phi)+3/2}}{\Phi}.
\end{equation} 
From now on, we will use the unit $\kappa=1$ for simplicity.

The equations of motion derived from action (\ref{EFA}) read 
\begin{eqnarray}
	G_{\mu\nu}^* = \kappa_*(T^{*\mathrm{m}}_{\mu\nu}+T_{\mu\nu}^{*\phi}), \label{cor1}\\
	\square_*\phi =\frac{\rmd V}{\rmd\phi}-\alpha T_\mathrm{m}^*,\label{cor2}
\end{eqnarray}
where $T^*_\mathrm{m}\equiv g^{\mu\nu}_*T^{*\mathrm{m}}_{\mu\nu}$ is the trace of
$T_{\mu\nu}^{*\mathrm{m}}$, and the energy-momentum tensors of the background fluid and
the scalar field are
\begin{eqnarray}
	T^{*\mathrm{m}}_{\mu\nu}=-\frac{2}{\sqrt{-g_*}}
	\frac{\delta(\sqrt{-g_*}\mathcal{L}_\mathrm{m})}{\delta
	g_*^{\mu\nu}},\label{cor3}\\
	T^{*\phi}_{\mu\nu}=\partial_\mu \phi \partial_\nu
	\phi-g^*_{\mu\nu}\left[\frac12g^{\alpha\beta}_* \partial_\alpha
	\phi\partial_\beta\phi+V(\phi)\right].\label{cor4}
\end{eqnarray}
When $\alpha = 0$, the field equations of scalar-tensor gravity in the Einstein
frame, Equations \eref{cor1} and \eref{cor2}, along with energy-momentum
tensors in Equations \eref{cor3} and \eref{cor4}, are in the same form of the
field equations \eref{GRQ} and \eref{eqphi} and the energy-momentum tensors
in Equations \eref{T} and \eref{TQ} for the quintessence model, respectively,
so that the gravity is entirely described by the metric tensor. The modification of
gravity comes from the direct coupling term $\alpha(\phi) T_\mathrm{m}^*$ in \Eref{cor2} between the fluid and the scalar field $\phi$. Hence, the local
conservation of the energy-momentum tensor no longer holds in the Einstein
frame, and it changes to
\begin{equation}\label{ff}
	\nabla^*_\mu T^{\mu\nu}_{*\mathrm{m}}=\alpha(\phi)T_\mathrm{m}^*\nabla^\nu_*\phi.
\end{equation}
As a result, the test particle does not move along geodesics of $g^*_{\mu\nu}$
anymore, as if it suffers from a kind of ``fifth force". According to \Eref{ff}, the original quintessence model is recovered as $\alpha(\phi)$
vanishes, which corresponds to the case where the function $W(\Phi)$ in the
Jordan frame diverges. In this subsection we discuss the results in the Einstein
frame, and the star notation of physical quantity is omitted in the following
part of this subsection for simplicity. But keep in mind that we are in the
Einstein frame.

\subsubsection{Dynamics of Coupled Quintessence.}\label{s421}

The fact that the energy density of dark energy is of the same order as that of
dark matter in the present Universe suggests that there may be some relation
between them. Among various coupled dark energy models \cite{Tsujikawa:2010sc}, the
coupled quintessence includes an interaction between a scalar field $\phi$ and
the dust  matter---nonrelativistic matter whose pressure is zero---in the form
\cite{Wetterich:1994bg,Amendola:1999er},
\begin{equation}\label{cd}
	\nabla^\mu T_{\mu\nu}^\phi=-\alpha(\phi)T_\mathrm{d}\nabla^\nu\phi,\quad\nabla^\mu
	T^\mathrm{d}_{\mu\nu}=\alpha(\phi)T_\mathrm{d}\nabla^\nu\phi, 
\end{equation}
where $T_{\mu\nu}^\phi$ and $T^\mathrm{d}_{\mu\nu}$ are the energy-momentum tensors of
the scalar field and the dust. The coupling of the scalar field and
radiation vanishes due to the vanishing trace of $T_{\mu\nu}^\mathrm{r}$, the
energy-momentum tensor of the radiation. Generally speaking, the coupling
strength $\alpha(\phi)$ between the scalar field and various components of the
Universe is not necessarily uniform
\cite{Damour:1990tw,Casas:1991ky,Holden:1999hm,Fuzfa:2006pn}, and the baryons
are usually treated uncoupled to the scalar field to avoid an extra long-range
force besides gravity.

We only consider the case where the coupling $\alpha$ is a constant and the
potential is exponential for simplicity,\footnote{In the general scalar-tensor
theory with action \eref{EQAA}, the constant coupling condition is
approximately satisfied, according to \Eref{Q}, if
\begin{equation}
	\left(\frac{\rmd F}{\rmd\varphi}\right)^2\gg\frac23\zeta F.
\end{equation}
Then, one gets the function $F(\phi)=\exp(\pm\sqrt{2/3}\phi)$ and the constant
coupling $\alpha\approx \pm1/\sqrt{6}$. Furthermore, a simple way to construct
the exponential potential $V(\phi)$ is assuming that the coupling function
$F(\varphi)$ and the potential function $U(\varphi)$ satisfy the relation
\begin{equation}\label{UV}
	U(\varphi)\propto F^k(\varphi).
\end{equation}
which holds, for instance, when both $U$ and $F$ are power-law or exponential,
but is also valid for much more complicated functions, like products of
power-law and exponential forms. The relation \eref{UV} was first used by
Amendola et al. \cite{Amendola:1993it} to reveal the relation between the
scalar-tensor theory and the coupled quintessence, and to study the non-minimal
coupling gravity in the strong coupling limit. However, the solutions of cosmic
evolution are ruled out by the present constraints on the variability of the
gravitational coupling, and they only allow for an energy density $\Omega_\phi \simeq 0.04$ in the form of quintessence \cite{Amendola:1999qq}.}
then the Friedmann equations are Equations \eref{QRW} and \eref{QRW'}, which
are the same as for the quintessence. However, the evolution of the dust and the
scalar field obeys different equations
\begin{eqnarray}
	\dot\rho_\mathrm{d}+3H\rho_\mathrm{d}=\alpha\rho_\mathrm{m}\dot\phi,\\
	\dot\rho_\phi+3H(\rho_\phi+p_\phi)=-\alpha\rho_\mathrm{m}\dot\phi,
\end{eqnarray}
where $\rho_\phi$ and $p_\phi$ have already been shown in Equations \eref{rhoQ}
and \eref{pQ}.

\begin{figure}[h]
\centering
\begin{minipage}{0.6\textwidth}
  \centering
  \includegraphics[width=\linewidth]{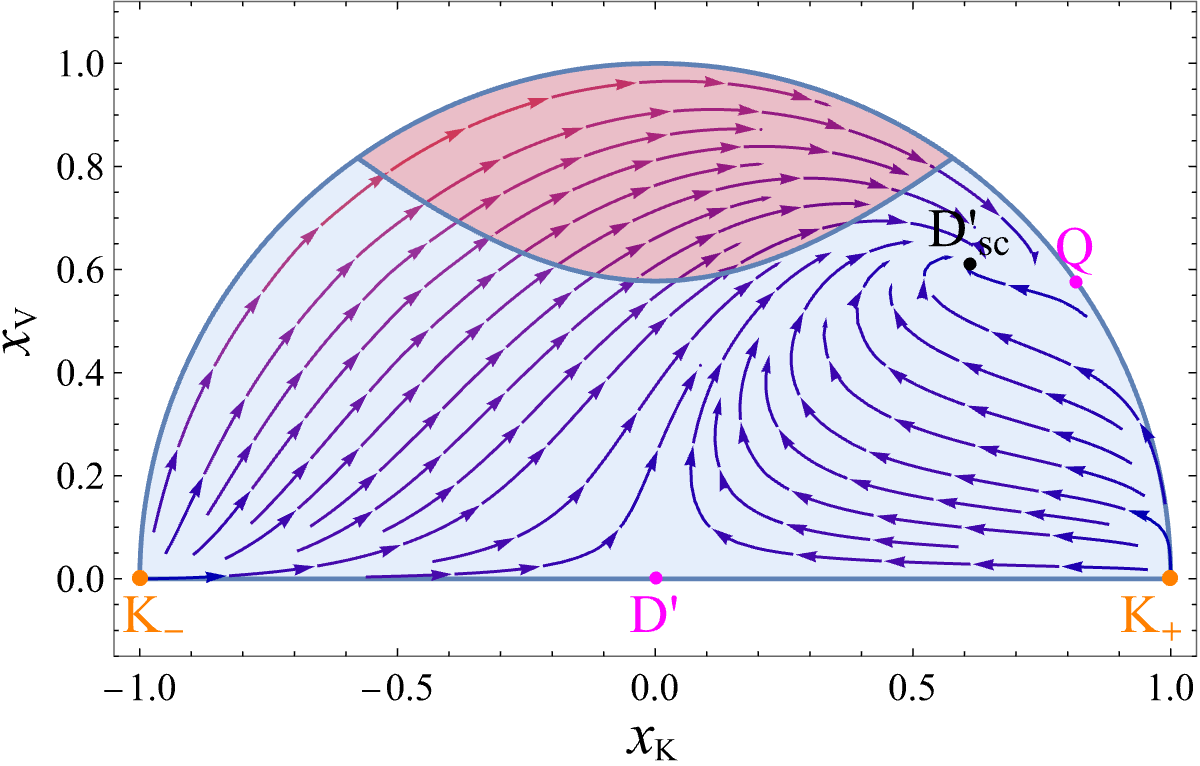}
\end{minipage}
\begin{minipage}{0.6\textwidth}
  \centering
  \includegraphics[width=\linewidth]{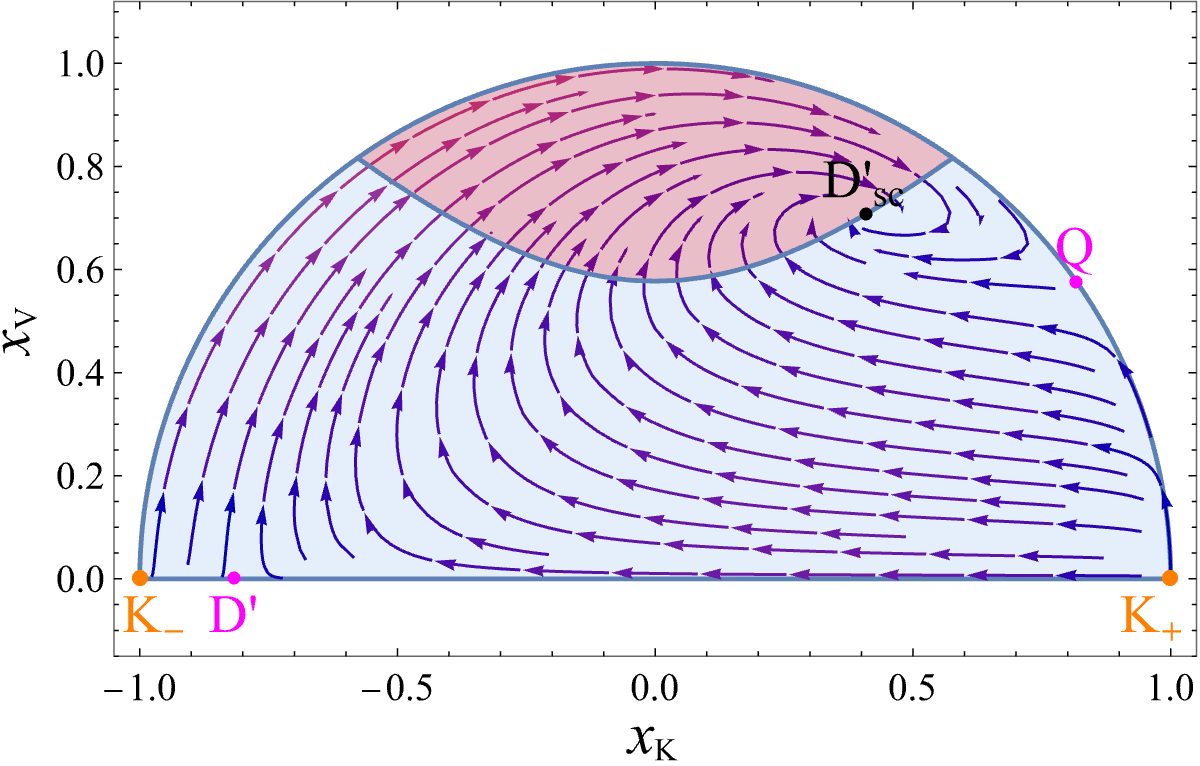}
\end{minipage}
\begin{minipage}{0.6\textwidth}
  \centering
  \includegraphics[width=\linewidth]{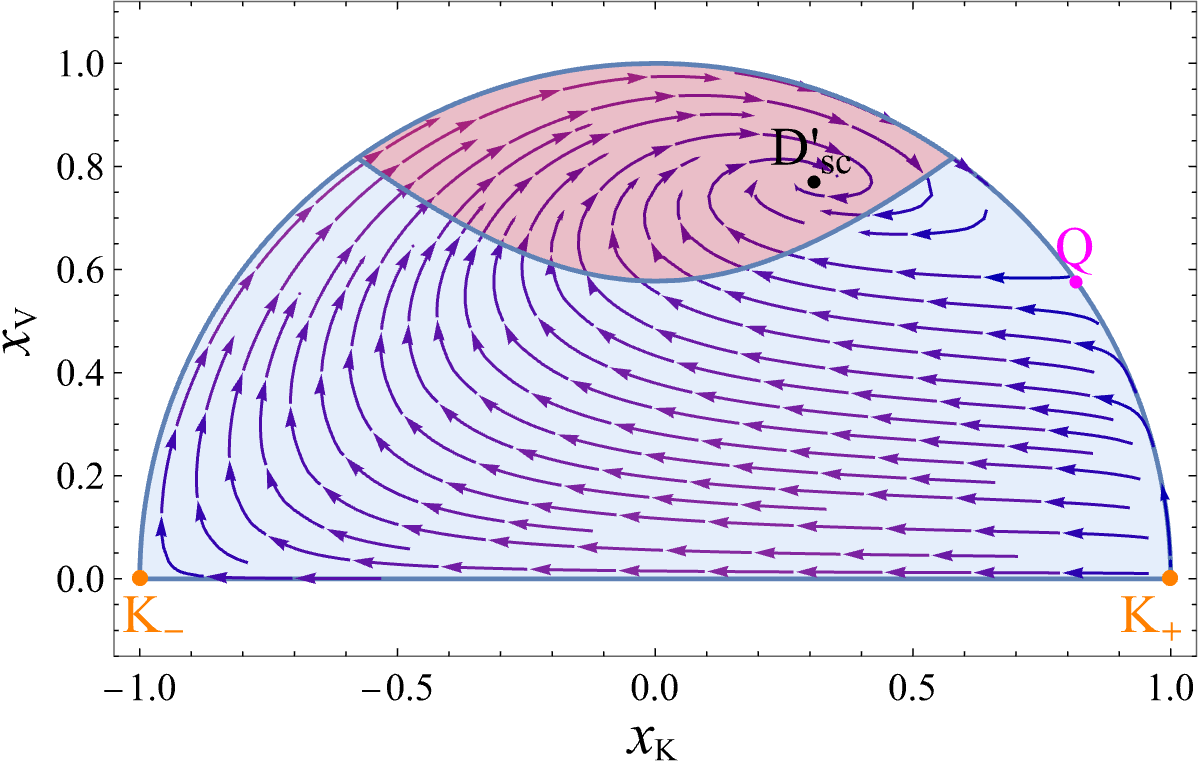}
\end{minipage}
\caption{The phase space of $x_\mathrm{K}$ and $x_\mathrm{V}$ with $\alpha=0$ (the upper subfigure), $\alpha=1$ (the middle subfigure), and $\alpha=2$ (the bottom subfigure). Here we assume $x_\mathrm{r}=0$ and $\lambda=1$. The fixed points of coupled quintessence are plotted, with yellow, magenta, and black representing stable, saddle, and unstable fixed points, respectively. Each streamline represents the evolution of the Universe under different initial conditions. The red region identifies accelerated expansion. Each point on the curve $x_\mathrm{K}^2+x_\mathrm{V}^2=1$ corresponds to $\Omega_\mathrm{q}=1$. Since the future attractor $\mathrm{D}_\mathrm{sc}'$ in the bottom subfigure lies in the red region but not on the semi-circle, it offers a solution to the coincidence problem.}
\label{fig:cq}
\end{figure}

Introducing the same expansion normalised variables $x_\mathrm{K}$, $x_\mathrm{V}$ and $x_\mathrm{r}$ defined in \Eref{ds}, we
have an altered differential equation
\begin{equation}
	\frac{{\rmd} x_\mathrm{K}}{\rmd N}=-3 x_\mathrm{K}+\frac{\sqrt{6}}{2} \lambda x_\mathrm{V}^2-x_\mathrm{K} \frac{1}{H}
	\frac{\rmd H}{\rmd N}-\frac{\sqrt{6}}{2} \alpha(1-x_\mathrm{K}^2-x_\mathrm{V}^2-x^2_\mathrm{r})
\end{equation}
for $x_\mathrm{K}$, while the equations for $x_\mathrm{V}$ and $x_\mathrm{r}$ do not change, so as the EOS
parameters and density parameters defined in \Sref{ADS}. Except for the
fixed points in \Tref{table1}, additional ones of the coupled quintessence
are listed in \Tref{table1.5}, and \Fref{fig:cq} shows fixed points in the phase space with $\lambda=1$ and different coupling constant $\alpha$.

\begin{table}
\centering
\caption{Properties of additional fixed points for the coupled quintessence with
a constant $\alpha$ and potential \eref{V} \cite{Amendola:1999er,amendola2010dark}.}\label{table1.5}
    \begin{tabular}{@{}lllllll}
    \br
Fixed point	& $(x_\mathrm{K},x_\mathrm{V},x_\mathrm{r})$ &$\Omega_\mathrm{r}$ & $\Omega_\mathrm{d}$& $\Omega_\phi$   &   $w_\phi$
&   $w_\mathrm{eff}$\\
\mr
$\mathrm R_\mathrm{ex}$	&
$\Big(-\frac{1}{\sqrt{6}\alpha},0,\frac{\sqrt{2\alpha^2-1}}{\sqrt{2}\alpha}\Big)$
& $1-\frac{1}{2\alpha^2}$&$\frac{1}{3\alpha^2}$ & $\frac{1}{6\alpha^2}$ & $1$ &
$\frac{1}{3}$\\
$\mathrm D'$  &  $\Big(-\frac{\sqrt6 \alpha}{3},0,0\Big)$	&   $0$
&$1-\frac{2\alpha^2}{3}$ &$\frac{2\alpha^2}{3}$ &   $1$ &
$\frac{2\alpha^2}{3}$\\
$\mathrm D_\mathrm{sc}'$   &
$\Big(\frac{\sqrt{6}}{2(\alpha+\lambda)},
\frac{\sqrt{2\alpha^2+2\alpha\lambda+3}}{2(\alpha+\lambda)},0\Big)$ & $0$
&$\frac{\lambda^2+\alpha\lambda-3}{(\alpha+\lambda)^2}$&
$\frac{\alpha^2+\alpha\lambda+3}{(q+\lambda)^2}$  &
$-\frac{\alpha(\alpha+\lambda)}{\alpha(\alpha+\lambda)+3}$   &
$-\frac{\alpha}{\alpha+\lambda}$\\
\br
\end{tabular}
\end{table}

In the coupled quintessence model, fixed point $\mathrm D'$ ($\mathrm
D_\mathrm{sc}'$) replaces D ($\mathrm D_\mathrm{sc}$) in the uncoupled case, and
$\mathrm D'$ ($\mathrm D_\mathrm{sc}'$) returns to D ($\mathrm D_\mathrm{sc}$) when
the coupling $\alpha$ vanishes. Notice that both $\mathrm D'$ and $\mathrm
D_\mathrm{sc}'$ are scaling solutions, but $\mathrm D'$ returns to a domination
solution while $\mathrm D_\mathrm{sc}'$ returns to a scaling solution as
$\alpha\to0$. The fixed point $\mathrm R_\mathrm{ex}$ represents an extra phase of the
radiation domination.

Two fixed points, Q and $\mathrm D_\mathrm{sc}'$, are possible to represent the
present accelerating phase of the Universe, giving two feasible evolutions. The
first is the sequence $\mathrm R\to\mathrm D'\to \mathrm D_\mathrm{sc}'$, which is
able to give rise to a global attractor with $\Omega_d\sim0.3$ and
$\Omega_q\sim0.7$, so that it can be used for alleviating the coincidence
problem. However, the coupled quintessence with an exponential potential does
not allow for such a cosmological evolution, because the condition
$\alpha^2\ll1$ is required to have point $\mathrm D'$ compatible with
observations whereas large values of $|\alpha|$ are needed to get the late-time
cosmic acceleration \cite{Amendola:1999er}. A solution to this problem is to
consider a step-like function of the coupling $\alpha$ \cite{Amendola:2000uh}.

The other evolution route is $\mathrm R\to\mathrm D'\to \mathrm Q$, where the
phase that is intermediate between the radiation dominated epoch and the accelerated epoch is
$\mathrm D'$, a saddle point in the phase space which replaces point D in the
uncoupled model. The presence of the $\phi$-matter domination era changes the
background expansion history of the Universe as $a\propto t^{2/(3+2\alpha^2)}$.
Therefore, one gets a smaller sound horizon at the decoupling epoch, as well as
a larger growth rate of matter perturbations, relative to the uncoupled
quintessence. According to them, the CMB data and the Lyman-$\alpha$ power
spectra would put an upper bound $\alpha\sim0.1$ on the coupling strength
\cite{Amendola:2003eq,DiPorto:2007ovd}.

\subsubsection{Chameleon Mechanism.}

Although the general couplings between background fluid and scalar field are not
necessarily universal, the preferential coupling of dark energy to dark matter
looks like a ``conceptual fine-tuning". If the scalar field couples to baryons,
unless the coupling is weak enough, a long-range fifth force should be observed.
Nevertheless, the experimental bounds on the coupling would not effectively
apply on the large cosmological scale if some kind of screening mechanism works.

The chameleon mechanism \cite{Khoury:2003aq,Khoury:2013yya} effectively shields
the fifth force by mediating the dynamics of the scalar field with the matter
density of the environment. Consequently, the behavior of the scalar field
varies in different environments, analogous to a chameleon adjusting its color
in response to various surroundings.

Considering the chameleon mechanism, the motion equation of the scalar field
\eref{cor2} is rewritten as
\begin{equation}\label{KGeff}
	\square\phi=\frac{\rmd V}{\rmd\phi}-\alpha \rme^{\alpha\phi}\hat\rho_\mathrm{d},
\end{equation}
where $\hat\rho_\mathrm{d}\equiv \rme^{-\alpha\phi}\rho_\mathrm{d}$, satisfying
\begin{equation}
	\dot{\hat\rho}_\mathrm{d}+3H\hat\rho_\mathrm{d}=0,
\end{equation}
thus $\hat\rho_\mathrm{d}$ is conserved in the Einstein frame. A non-trivial assumption
of the chameleon mechanism is that the matter density considered in the
Klein-Gordon equation of motion  is the conserved $\hat\rho_\mathrm{d}$ in the Einstein
frame, which is independent of $\phi$. According to \Eref{KGeff}, the
dynamics of the chameleon is not governed  by $V(\phi)$ alone, but by the
effective potential,
\begin{equation}
	V_\mathrm{eff}(\phi,\hat\rho_\mathrm{d})=V(\phi)+\hat\rho_\mathrm{d}\rme^{\alpha\phi},
\end{equation}
which is a function also depends on the dust
density of the environment. 

The value of $\phi$ at the minimal of the potential, $\phi_\mathrm{min}$, and the mass squared of small fluctuations around the minimum, $m_\mathrm{eff}^2=\partial_\phi^2V_\mathrm{eff}(\phi_\mathrm{min})$ depend on the environment as well, provided such a minimum exists. The denser the environment, the more massive the chameleon is. The effective mass, in turn, determines the reach of the Yukawa-type potential for the interaction, in a form $\propto \rme^{-m_\mathrm{eff}r}/r$. The larger the effective mass is, the weaker the fifth-force associated with the chameleon field is, since it results in a faster decay of the interaction with the distance. For a massive object, the fifth force from deep within to the exterior profile is Yukawa-suppressed. Consequently, only the contribution from within a thin shell beneath the surface significantly affects the exterior profile, a phenomenon known as the thin-shell effect. Since the chameleon effectively couples only to the shell, whereas gravity of course couples to the entire bulk of the object, the chameleon force on an exterior test mass is suppressed compared to the gravitational force. To satisfy solar system tests, the Milky Way galaxy must be screened, which gives the condition \cite{Khoury:2013yya}
\begin{equation}
    \ln\frac{A(\phi_0)}{A(\phi_\mathrm{MW})}\lesssim10^{-6}.
\end{equation}
Here, $\phi_0$ represents the cosmic background scalar field today, while $\phi_\mathrm{MW}$ denotes the ambient field value at the center of the Milky Way, and $A(\phi)$ is the function in the conformal transformation \eref{CT}.

The prototypical chameleon potential considers the Ratra-Peebles potential \eref{RPV}, thereby the effective potential is given by, up to an irrelevant constant,
\begin{equation}
    V_\mathrm{eff}(\phi)=\frac{M^{4+n}}{\phi^4}+\alpha\hat{\rho}_\mathrm{d}\phi,
\end{equation}
where uses the fact that $\phi\ll1$ over the relevant field range. For a positive coupling $\alpha$, it displays a minimum at $\phi_\mathrm{min}\propto\hat\rho_\mathrm{d}^{-1/(n+1)}$ and an effective mass $m_\mathrm{eff}\propto\hat\rho_\mathrm{d}^{(n+2)/(n+1)}$ that increases when the ambient matter gets denser.

The most stringent constraint on this model arises from laboratory tests of the inverse-square law, which impose an upper limit of approximately 50 $\mathrm{\mu m}$ on the range of the fifth force, assuming gravitational strength coupling \cite{Adelberger:2006dh}. The constraints on deviations from general relativity, including effective principles as well as post-Newtonian tests in the solar system and observations of binary pulsars, translate to an upper limit on $M$ of around $10^{-3}\;\mathrm{eV}$, which coincides with the dark energy scale \cite{Khoury:2003aq,Khoury:2003rn}.

An intriguing observation is that the vacuum expectation value of the chameleon field increases in high-density environments. In other words, the effective cosmological constant also increases, consequently leading to a higher local Hubble expansion rate. This suggests the chameleon dark energy model could potentially serve as a model to elucidate the Hubble tension (and even the $S_8$ tension) \cite{Cai:2021wgv}.

\subsection{Jordan Frame: Extended Quintessence}\label{subsec2}

Quintessence modeled by a non-minimally coupled scalar field is called the
extended quintessence. The action for the extended quintessence is given by
Perrotta et al. \cite{Perrotta:1999am},
\begin{equation}\label{eQAqqq}
	\fl S[g^{\mu\nu},\varphi,\Psi_\mathrm{m}]=\int \rmd^4 x \sqrt{-g}\left[\frac{1}{2}
	f(\varphi, R)-\frac{1}{2} \zeta(\varphi)g^{\mu\nu}\partial_\mu\varphi\partial_\nu\varphi\right]+ S_\mathrm{m}[g^{\mu
	\nu}, \Psi_\mathrm{m}],
\end{equation}
where $f$ is a general function of the scalar field $\varphi$ and the Ricci
scalar $R$, and $\zeta$ is a function of $\varphi$ only.  The action
\eref{eQAqqq} includes a wide variety of theories such as the $f(R)$ gravity \cite{DeFelice:2010aj,Capozziello:2011et},
where
\begin{equation}
	f(\varphi,R)=f(R),\quad\zeta(\varphi)=0,
\end{equation}
the BD gravity, where
\begin{equation}
	f(\varphi,R)=\varphi
	R,\quad\zeta(\phi)=\frac{\omega_{\mathrm{BD}}}{\varphi},
\end{equation}
and the dilaton gravity \cite{Overduin:1997sri}, where
\begin{equation}\label{dg}
	f(\varphi,R)=2\rme^{-\varphi}R-2U(\varphi),\quad\zeta(\varphi)=-2\rme^{-\varphi}.
\end{equation}

Here we focus on particular cases when $f(\varphi,R)=F(\varphi)R-2U(\varphi)$,
and the action becomes
\begin{equation}\label{EQAA}
	\fl S[g^{\mu\nu},\varphi,\Psi_\mathrm{m}]=\int
	\rmd^4x\sqrt{-g}\left[\frac{1}{2}F(\varphi)R-\frac12\zeta(\varphi)g^{\mu\nu}\partial_\mu\varphi\partial_\nu\varphi
	-U(\varphi)\right]+S_\mathrm{m}[g^{\mu\nu},\Psi_\mathrm{m}].
\end{equation}
The related equations of motion are
\begin{eqnarray}
\fl FG_{\mu \nu}=T_{\mu \nu}^{\mathrm{m}}  +\zeta\partial_\mu \varphi
\partial_\nu \varphi-\frac{\zeta}{2} g_{\mu
\nu}g^{\alpha\beta}\partial_\alpha\varphi\partial_\beta\varphi-g_{\mu
\nu}U +(\nabla_\mu \nabla_\nu-g_{\mu \nu} \square) F,\\
\fl \zeta \square \varphi+\partial_\varphi\zeta \cdot g^{\mu\nu}\partial_\mu\varphi\partial_\nu\varphi
+\frac{\partial_\varphi F}{2} R=\partial_\varphi U.\label{K-G}
\end{eqnarray}

\subsubsection{Non-minimal Coupling Theory.}

In order to illustrate the dynamics of the extended quintessence scenario, we consider the non-minimal coupling theroy,
\begin{equation}\label{fz}
	F(\varphi)=1-\xi\varphi^2,\quad\zeta(\varphi)=1,
\end{equation}
where $\xi$ is the non-minimal coupling constant. Special values of the constant
$\xi$ have received particular attention in literature. For example, $\xi=1/6$
corresponds to the conformal coupling, because the Klein-Gordon equation
(\ref{K-G}) and the physics of $\varphi$ are conformally invariant if $U = 0$ or
$U \propto\varphi^4$ \cite{Wald:1984rg}. Besides that, $\xi=0$ and $|\xi|\gg1$
are the minimal coupling and strong coupling, respectively.

One defines the energy-momentum tensor of the scalar field as\footnote{There are
different possible definitions of the effective energy-momentum tensor for the
scalar field in scalar-tensor theories, and the effective energy density,
pressure, and EOS are unavoidably linked to one of these definitions. See
Ref.~\cite{Bellucci:2001cc} for non-minimal coupling theory and
Ref.~\cite{Torres:2002pe} for the BD theory.} \cite{Uzan:1999ch}
\begin{equation}
	\fl T_{\mu\nu}^\varphi = \partial_\mu\varphi\partial_\nu\varphi -g_{\mu\nu}\left[\frac12g^{\alpha\beta}\partial_\alpha\varphi\partial_\beta\varphi+U(\varphi)\right] + 
	\xi(g_{\mu\nu}\square-\nabla_\mu\nabla_\nu)(\varphi^2)
	+ \xi\varphi^2G_{\mu\nu}.
\end{equation}
The field equation of the tensor field becomes
\begin{equation}\label{EQGG}
	G_{\mu\nu}=T_{\mu\nu}^{\mathrm{m}}+T_{\mu\nu}^\varphi.
\end{equation}
Then the cosmological dynamics is governed by
\begin{eqnarray}
	3H^2=\rho_\mathrm{m}+\rho_\varphi,\\
	-2\dot H=\rho_\mathrm{m}+\rho_\varphi+p_m+p_\varphi ,
\end{eqnarray}
where
\begin{eqnarray}
	\rho_\varphi =\frac{1}{2}\dot\varphi^2 +V
	+3H\xi\varphi(2\dot\varphi+H\varphi),\\
	p_\varphi =\frac{1}{2}\dot\varphi^2-V - \xi\Big[(2\dot H+3H^2)\varphi^2 +
	4H\varphi\dot\varphi+2\varphi\ddot\varphi+2\dot\varphi^2\Big].
\end{eqnarray}
The  equation of motion for the scalar field in a flat FLRW background is given by
\begin{equation}
	\ddot\varphi+3H\dot\varphi+\frac{\rmd U_\mathrm{eff}}{\rmd\varphi}=0,
\end{equation}
where 
\begin{equation}
	U_\mathrm{eff}=U+\frac{\xi}{2}R\varphi^2 \,,
\end{equation}
and	 $R=6(2H^2+\dot H)$ is the Ricci curvature scalar. At a sufficiently early
time when the curvature is high enough, the non-minimal coupling term dominates
over the self-interaction potential $U$. Then the field $\varphi$ settles down
to a slow-roll regime where the friction term, $3H\dot\varphi$, balances the
term $\xi R\varphi$. After that $\varphi$ starts to roll fast, then the coupling
can be ignored and the field behaves as a minimally coupled field.

To have an accelerated expansion with $0<\xi<1/6$, one needs a potential
$U(\varphi)$ that does not grow with $\varphi$ faster than the function
\begin{equation}\label{C}
	C(\varphi)\equiv\bar U\varphi^k \exp\left(\frac{\varphi^2}{6}\right),
\end{equation}
where $\bar U$ is a constant, and $k\approx0.26/\xi$ by assuming $\Omega_m
\simeq 0.3$ and $\Omega_\varphi \simeq 0.7$ today \cite{Faraoni:2004pi}. If
instead $\xi<0$, then $U(\varphi)$ must grow faster than $C(\varphi)$. In this
case, $k$ is negative and, upon a rescaling of $\varphi$, the function \eref{C}
reduces to the supergravity potential \cite{Brax:1999gp}.

When the Ratra-Peebles potential is used with $\xi\neq0$, the tracking solution
[see \Eref{wQ}]
\begin{equation}\label{EQTS}
	w_\varphi=\frac{nw_\mathrm{m}-2}{n+2}
\end{equation}
is still valid, where $n$ is the parameter in the Ratra-Peebles potential
(\ref{RPV}). This shows that the scaling solution does not depend on the
coupling $\xi$, and the solution is always stable since the value of $\xi$ only
determines the nature of the stable point \cite{Uzan:1999ch}. This relation
\eref{EQTS} generalizes the one found for minimally coupled scalar fields
\cite{Ratra:1987rm,Zlatev:1998tr,Liddle:1998xm}.

The value of $\xi$ however, cannot be arbitrary because the scalar field is
ultra-light. It mediates a long-range force that is constrained by Solar system
experiments. From effects induced on photon trajectories \cite{Will:2014kxa} and \Eref{BDnm}, one has
\begin{equation}\label{res1}
	\omega_\mathrm{BD}=\frac{F_0}{F_0'^2}>4\times10^4,
\end{equation}
which gives
\begin{equation}
	|\xi|<\frac{4.3\times10^{-3}}{\sqrt{n(n+2)}}.
\end{equation}
Here we write the abstract function $F(\varphi)$ instead of a specific shape because the restrictions
\eref{res1}, as well as \Eref{res2} below, are generally valid for
any form of $F$. Also, a constraint from the time variation of the gravitational
constant \cite{Wu:2009zb}
\begin{equation}\label{res2}
	\left.\frac{\dot G_\mathrm{eff}}{G_\mathrm{eff}}\right|_0=\frac{\dot
	F_0}{F_0}\leq10^{-11}\;{\rm yr}^{-1}
\end{equation}
yields a constraint on $\xi$ \cite{Chiba:1999wt}
\begin{equation}
	-10^{-2}\lesssim\xi\lesssim10^{-1}
\end{equation}
for the Ratra-Peebles potential.

The scalar field, as an additional source of fluctuations, can cause new and
observable effects in the CMB and in the formation of LSSs
\cite{Viana:1997mt}. In addition, the time variation of the potential
$U_\mathrm{eff}$ between the last scattering surface and the present time would
enhance the integrated Sachs-Wolfe effect \cite{Perrotta:1998vf}. The
non-minimal coupling in extended quintessence models also modifies the positions
of the acoustic peak multipoles, which can be $\sim10\%$ to $\sim30\%$ with
respect to the standard quintessence \cite{Perrotta:1999am}.

\subsubsection{The Brans-Dicke theory.}\label{br-di}

Before going through the general scalar-tensor theory, it is helpful to analyze a simple case---the Brans-Dicke theory---to acquire an advance impression of how non-minimally coupling diverges from the minimal approach in cosmic evolution.
Usually, to meet the constraint \eref{res1}, a non-self-interacting BD scalar
cannot be in the form of quintessence \cite{Faraoni:2004pi}, and one needs a BD
theory with a potential,
\begin{equation}\label{UBDA}
	\fl S[g^{\mu\nu},\Phi,\Psi_\mathrm{m}]=\int \rmd^4x\sqrt{-g} \left[\frac {\Phi}2R
	-\frac{\omega_\mathrm{BD}}{2\Phi} g^{\mu\nu}\partial_\mu
	\Phi\partial_\nu \Phi-U(\Phi)\right]+ \int \rmd^4x\sqrt{-g}\mathcal
	L_\mathrm{m}.
\end{equation}
Notice that the action \eref{UBDA} is equivalent to the action of a non-minimal
coupling theory by applying a redefinition of the scalar field $\Phi\to
F(\varphi)$ that satisfies
\begin{equation}\label{BDnm}
	F\left(\frac{\rmd F}{\rmd\varphi}\right)^{-2}={\omega_\mathrm{BD}}.
\end{equation}

To avoid the obvious fraction term in the action, we need the so-called string frame representation
\cite{Lidsey:1999mc} of a dilatonic BD theory by rescaling the scalar field
$\Phi=\rme^\psi$, into
\begin{equation}\label{SFA}
	\fl S[g^{\mu\nu},\psi,\Psi_\mathrm{m}] =\int \rmd^4x\sqrt{-g}\rme^\psi\left[\frac R{2}
	-\frac{\omega_\mathrm{BD}}2g^{\mu\nu}\partial_\mu\psi\partial_\nu\psi
	-V(\psi)+\rme^{-\psi}\mathcal L_\mathrm{m}\right],
\end{equation}
where $V(\psi)=U(\Phi)/\Phi$. Choosing 
\begin{equation}
y_\mathrm{K}\equiv\frac{\dot\psi}{\sqrt6 H},\quad x_\mathrm{V}\equiv\frac{\sqrt{V}}{\sqrt3 H},
\end{equation}
and $\lambda$ [defined in \Eref{lambda}] as variables of the phase space in the autonomous system, the
density parameters are defined as follows,
\begin{eqnarray}
    \Omega_\mathrm{K}^\mathrm{eff}\equiv w_\mathrm{BD}y_\mathrm{K}^2-\sqrt6y_\mathrm{K},\\
	\Omega_\mathrm{V}\equiv x_\mathrm{V}^2,\\
	\Omega_\mathrm{m}^\mathrm{eff}\equiv\frac{\rme^{-\psi}\rho_\mathrm{m}}{3H^2}.
\end{eqnarray}
From the mathematical expression, the effective kinetic energy density,
$\Omega^\mathrm{eff}_\mathrm{K}$---unlike $\Omega_\mathrm{K}$ in the quintessence model---has not to
be positive, and it is rational in physics because the scalar field is a part of
gravity in a modified theory rather than a part of cosmic contents.

To investigate the asymptotic dynamics of the BD cosmological model, it is
useful to consider the vacuum cosmology with $\Omega_\mathrm{m}^\mathrm{eff}=0$. The
autonomous ordinary differential equations of BD theory are derived from the
equations of motion
\cite{Garcia-Salcedo:2015naa},
\numparts
\begin{eqnarray}
		\frac{\rmd y_\mathrm{K}}{\rmd N} =(\omega_{\mathrm{BD}} y_\mathrm{K}^2-\sqrt{6}
		y_\mathrm{K}-1)\left[3y_\mathrm{K}-\frac{(1+\lambda)(3y_\mathrm{K}
		+\sqrt6)}{2\omega_\mathrm{BD}+3}\right], \\
		\frac{\rmd\lambda}{\rmd N}=-\sqrt{6} \lambda^2(\Gamma-1)y_\mathrm{K},
\end{eqnarray}
\endnumparts
where $N=\ln a$ is the $\mathrm{e}$-folding number and $\Gamma$---assumed to be written as
a function of $\lambda$---is defined in \Eref{gamma}. We only consider
a non-negative potential $V(\psi)$, i.e. $\Omega_\mathrm{V}\ge0$, which restricts the
range of $y_\mathrm{K}$ in the phase space,
\begin{equation}
	y_-\le y_\mathrm{K}\le y_+,\quad y_\pm =
	\frac{\sqrt{3}\pm\sqrt{2\omega_\mathrm{BD}+3}}{\sqrt{2}\omega_\mathrm{BD}}.
\end{equation}
The bounds of $\lambda$, if any, are set by the concrete form of the potential.

\begin{table}
\caption{Properties of fixed points for the vacuum BD Universe \cite{Quiros:2019ktw}.}\label{table2}
\begin{indented}
\item[]\begin{tabular}{@{}llllll}
\br
Fixed point	& $y_\mathrm{K}$	& $\lambda$  &  $\Omega_\mathrm{V}$   &   $q$   &   $\Gamma$\\
\mr
GR-de Sitter  &  $0$	&   $-1$	&	 $1$	&   $-1$   &   $1$\\
BD-de Sitter &  $\frac1{\sqrt6(1+\omega_\mathrm{BD})}$	&   $0$	&
$\frac{12+17\omega_\mathrm{BD}+6\omega_\mathrm{BD}^2}{6(1+\omega_\mathrm{BD})^2}$	&
$-1$   &   ---\\
Stiff-dilaton  &  $y_\pm$	&   $0$	&	 $0$	&   $2+\sqrt{6}y_\pm$   &
---\\
\br
\end{tabular}
\end{indented}
\end{table}

The system has four fixed points listed in \Tref{table2}. The parameter $q$
in the table is the deceleration parameter.
For these solutions in the table, the GR-de Sitter solution is a
potential-dominated solution, which corresponds to the cosmological constant
domination in general relativity. In contrary, the BD-de Sitter phase, which
does not arise in the minimally coupled model, is a scaling solution, and it
arises even if the matter exists. The last two stiff-dilaton solutions are
dominated by the effective kinetic term.

Further investigation shows that this GR-de Sitter solution is stable. However,
it has been shown by Garcia-Salcedo et al. \cite{Garcia-Salcedo:2015naa} that
the BD cosmology does not have the $\Lambda$CDM phase as a universal attractor
unless the given potential approaches to the exponential form $V\propto \rme^\psi$
as an asymptote, since according to \Tref{table2}, the GR-de Sitter
solution corresponds to the exponential potential $V(\psi)\propto \rme^{\psi}$,
which amounts to the quadratic potential $U(\Phi)\propto \Phi^{2}$ in
terms of the standard BD field $\Phi$ in action \eref{UBDA}. It has also
been shown that very specific conditions on the coupling function $W(\Phi)$
are to be imposed for the given scalar-tensor gravity to have the GR-de Sitter
limit \cite{Billyard:1998kg,Barrow:2007ce}.

\subsubsection{Dynamics of the Extended Quintessence.}\label{412}
Now let us concentrate on the general scalar-tensor theory. To analyze the dynamics, it is convenient to use the scalar $\phi$ in the Einstein frame instead of the original $\varphi$. The suitable conformal transformation described by the parameter
\begin{equation}\label{Q}
	\alpha\equiv-\frac{\partial_\phi F}{2F}= -\frac{\partial_\varphi F}{2F}
	\left[\frac32\left(\frac{\partial_\varphi
	F}F\right)^2+\frac{\zeta}{F}\right]^{-1/2},
\end{equation}
takes the action \eref{EQAA} into 
\begin{eqnarray}\label{qA}
	\fl S[g^{\mu\nu},\phi,\Psi_\mathrm{m}]=&\int
	\rmd^4x\sqrt{-g}\left[\frac12F(\phi)R-\frac12(1-6\alpha^2)
	F(\phi)g^{\mu\nu}\partial_\mu\phi\partial_\nu\phi -U(\phi)\right]\nonumber\\&+S_\mathrm{m}[g^{\mu\nu},\Psi_\mathrm{m}].
\end{eqnarray}
The rescaled scalar field of the extended quintessence in the Einstein frame is
\begin{equation}\label{psi}
	\phi=\int \rmd\varphi\sqrt{\frac32\left(\frac{\partial_\varphi
	F}F\right)^2+\frac{\zeta}{F}}.
\end{equation}
In the uncoupled limit $\alpha\equiv\rmd\ln A/\rmd\phi\to 0$, the action \eref{qA} reduces to the action of quintessence model.

For simplicity, we treat $\alpha$ as a constant from now on,\footnote{Usually,
$\alpha$ is not a constant. For instance, the coupling $\alpha$ is
field-dependent in the non-minimal coupling theory,
\begin{equation}
	\alpha(\varphi)=\frac{\xi\varphi}{\sqrt{1-\xi(1-6\xi)\varphi^2}}.
\end{equation}
Here, $\alpha=\xi\varphi$ for $|\xi|\ll1$ and $\alpha=\pm1/\sqrt6$ in the strong
coupling limit $|\xi|\gg1$. Nevertheless, action \eref{qA} with a constant
$\alpha$ is able to represent some meaningful theories. The BD theory with a
potential in \Eref{UBDA} is equivalent to the case in which the
parameter $\omega_\mathrm{BD}$ is related to $\alpha$ via the relation
\begin{equation}\label{wBDQ}
	2\omega_\mathrm{BD}+3=\frac1{2\alpha^2}.
\end{equation}
In the general relativistic limit $\alpha\to0$, we have
$\omega_\mathrm{BD}\to\infty$ as expected. In addition, $f(R)$ theory is acquired
in the metric (Palatini) formalism when $\alpha=-1/\sqrt6$
($\alpha^2\to\infty$), and the dilaton gravity (\ref{dg}) is recovered when
$\alpha=1/\sqrt2$.} leading to
\begin{equation}
	F(\phi)=\rme^{-2\alpha\phi}.
\end{equation}
In a flat FLRW background, the variation of  action \eref{qA} with respect to
the metric $g_{\mu\nu}$ and the scalar field $\phi$ leads to
\begin{eqnarray}
3 F H^2=\frac{1}{2}\left(1-6 \alpha^2\right) F \dot{\phi}^2+U-3 H
\dot{F}+\rho_\mathrm{d}+\rho_\mathrm{r},\label{EQ1} \\
2 F \dot{H}=-\left(1-6 \alpha^2\right) F \dot{\phi}^2-\ddot{F}+H
\dot{F}-\rho_\mathrm{d}-\frac43\rho_\mathrm{r},\label{EQ2}\\
(1-6 \alpha^2) F\left(\ddot{\phi}+3 H \dot{\phi}+\frac{\dot{F}}{2F}
\dot{\phi}\right)+\frac{\rmd U}{\rmd\phi}+\alpha F R=0,\label{EQ3}
\end{eqnarray}
where the overdot denotes the derivative of the cosmic time $t$, and $\rho_\mathrm{d}$
($\rho_\mathrm{r}$) is the energy density of dust (radiation).  We introduce the
following variables of the autonomous system
\begin{equation}
	x_\mathrm{K}\equiv\frac{\dot\phi}{\sqrt6H},\quad y_\mathrm{V}\equiv
	\frac1H\sqrt{\frac{U}{3F}}, \quad
	y_\mathrm{r}\equiv\frac{1}{H}\sqrt{\frac{\rho_\mathrm{r}}{3F}}
\end{equation}
and density parameters for the scalar field, non-relativistic matter, and
radiation
\begin{equation}
\Omega_\phi\equiv(1-6\alpha^2)x_\mathrm{K}^2+y_\mathrm{V}^2+ 2\sqrt6\alpha
x_\mathrm{K},\quad\Omega_\mathrm{d}\equiv\frac{\rho_\mathrm{d}}{3FH^2},\quad\Omega_\mathrm{r}\equiv y_\mathrm{r}^2 .
\end{equation}
A similar relation
\begin{equation}
	\Omega_\phi+\Omega_\mathrm{d}+\Omega_\mathrm{r}=1
\end{equation}
comes from Equation (\ref{EQ1}). The deceleration parameter $q$ and the
effective EOS parameter of the Universe are
\begin{equation}
	q=-1-\frac{\dot H}{H^2},\quad w_\mathrm{eff}=-1-\frac23\frac{\dot H}{H^2},
\end{equation}
where
\begin{equation}
	\fl\frac{\dot H}{H^2}=-\frac{1-6\alpha^2}{2}(3+3x_\mathrm{K}^2 -3y_\mathrm{V}^2+
	y_\mathrm{r}^2-6\alpha^2x_\mathrm{K}^2 +2\sqrt{6}\alpha x_\mathrm{K})+3\alpha(\lambda y_\mathrm{V}^2-4\alpha).
\end{equation}
Using Equations \eref{EQ1} to \eref{EQ3}, one obtains the differential
equations for $x_\mathrm{K}$, $y_\mathrm{V}$ and $y_\mathrm{r}$ \cite{Tsujikawa:2008uc},
\numparts
\begin{eqnarray}
\fl\frac{\rmd x_\mathrm{K}}{\rmd N}= \frac{\sqrt{6}}{2}\left(\lambda y_\mathrm{V}^2-\sqrt{6}
x_\mathrm{K}\right)\nonumber
\\+\frac{\sqrt{6} \alpha}{2}\left[\left(5-6 \alpha^2\right) x_\mathrm{K}^2+2
\sqrt{6} \alpha x_\mathrm{K}-3 y_\mathrm{V}^2+y_\mathrm{r}^2-1\right]-x _\mathrm{K}\frac{\dot{H}}{H^2}, \label{ds1} \\
\fl\frac{\rmd y_\mathrm{V}}{\rmd N}= \frac{\sqrt{6}}{2}(2 \alpha-\lambda) x_\mathrm{K}
y_\mathrm{V}-y_\mathrm{V}\frac{\dot{H}}{H^2},\label{ds2}\\
\fl\frac{\rmd y_\mathrm{r}}{\rmd N}=\sqrt{6}\alpha x_\mathrm{K}y_\mathrm{r}-2y_\mathrm{r}-y_\mathrm{r}\frac{\dot{H}}{H^2},
\end{eqnarray}
\endnumparts
where $N=\ln a$ and $\lambda = -\partial_\phi U/U$ as above.

\begin{table}
\caption{Properties of fixed points for the extended quintessence with a constant $\alpha$ and potential \eref{V} in the absence of radiation \cite{Tsujikawa:2008uc}.}\label{table3}
\begin{indented}
\item[]\begin{tabular}{@{}llll}
\br
Fixed point	& $(x_K,y_V)$  &  $\Omega_d$   &   $w_\mathrm{eff}$   \\
\mr
$\mathrm D''$ &  $\Big(\frac{\sqrt{6}\alpha}{3(2\alpha^2-1)},0\Big)$	&
$\frac{3-2\alpha^2}{3(1-2\alpha^2)^2}$	&   $\frac{4\alpha^2}{3(1-2\alpha^2)}$ \\
$\mathrm{D_{sc}''}$  &
$\Big(\frac{\sqrt6}{2\lambda},\sqrt{\frac{3+2\alpha\lambda-6\alpha^2}{2\lambda^2}}\Big)$
& $1-\frac{3-12\alpha^2+7\alpha\lambda}{\lambda^2}$ &
$-\frac{2\alpha}{\lambda}$\\
$\mathrm{K}_\pm'$ &  $\Big(\frac1{\sqrt6(\alpha+1)},0\Big)$	&	 $0$	&
$\frac{3\mp\sqrt6\alpha}{3(1\pm\sqrt6\alpha)}$   \\
$\mathrm Q'$  &  $\Big(\frac{\sqrt{6}(4 \alpha- \lambda)}{6(4 \alpha^2 -\alpha
\lambda-1)},\sqrt{\frac{6-\lambda^2+8 \alpha \lambda-16 \alpha^2}{6(4
\alpha^2-\alpha \lambda-1)^2}}\Big)$  &  $0$  &  $-\frac{20 \alpha^2-9 \alpha
\lambda-3+\lambda^2}{3(4 \alpha^2-\alpha \lambda-1)}$\\
$\mathrm {Q_{dS}}$  &  $(0,1)$ & $0$ & $-1$\\
\br
\end{tabular}
\end{indented}
\end{table}

In the absence of radiation ($y_\mathrm{r}=0$), the fixed points of the system for a
constant $\lambda$ are listed in \Tref{table3}. The
dust-dominated epoch can be realized either by the point $\mathrm D''$ or by the
point $\mathrm{D_{sc}''}$. If the point $\mathrm D''$ is responsible for the
matter domination, the condition $\alpha^2\ll1$ is required, leading to
$\Omega_\mathrm{m}\simeq1+10\alpha^2/3>0$ and $w_\mathrm{eff}\simeq{4\alpha^2}/{3}$. When
$\alpha^2\ll1$, the scalar-field dominated point $\mathrm{Q}'$ yields an
accelerated expansion of the Universe provided that
$-\sqrt2+4\alpha<\lambda<\sqrt2+4\alpha$ \cite{Tsujikawa:2008uc}. The scaling solution
$\mathrm{D_{sc}''}$ can give rise to the EOS, $w_\mathrm{eff}\simeq 0$ for
$|\alpha|\ll|\lambda|$. In this case, however, the condition $w_\mathrm{eff}<-1/3$
for the point $\mathrm{Q}'$ gives $\lambda^2<2$. Then the energy fraction of the
matter for the point $\mathrm{D_{sc}''}$ does not satisfy the condition
$\Omega_\mathrm{m}\simeq1$. As a result, the only way to describe the evolution from
matter domination to scalar-field domination in the phase diagram is the trace
from point $\mathrm{D}''$ to point $\mathrm{Q}'$. Note that fixed points in
\Tref{table3} tend to become the corresponding fixed points in
\Tref{table1} as $\alpha\to 0$ in the limit of general relativity.

Not surprisingly, the GR-de Sitter phase arises in the extended quintessence
model. If $\lambda=4\alpha$, the fixed point $\mathrm{Q}'$ turns into GR-de Sitter point
$\mathrm {Q_{dS}}$, with which corresponds $\Omega_\mathrm{d}=0$ and $w_\mathrm{eff}=-1$. Similar to the mentioned
quintessence model in \Sref{ADS}, we have to consider the
varying-$\lambda$ case in order to make it possible for the Universe to evolve
from the scaling solution $\mathrm{D_{sc}''}$ to the acceleration phase
$\mathrm{Q}'$ or $\mathrm {Q_{dS}}$. One of feasible potentials has been
found by Tsujikawa et al. \cite{Tsujikawa:2008uc}.

The dynamics of the extended quintessence system is much richer than in the
minimally coupled case \cite{foster1998scalar,Gunzig:2000ce,Gunzig:2000kk,Torres:2002pe,Hrycyna:2008gk}.  One can have spontaneous bouncing, spontaneous
entry into and exit from inflation, and super-acceleration ($\dot H>0$) in
extended models \cite{Gunzig:2000kk}. The phantom EOS and the boundary crossing of cosmological constants naturally arise in scalar-tensor theories with large
couplings \cite{Tsujikawa:2008uc}. Even in the limit $|\alpha|\ll1$, the phantom
EOS can be realized without introducing a ghost field
\cite{Perivolaropoulos:2005yv,Nesseris:2006jc,Martin:2005bp,Gannouji:2006jm}.

\subsubsection{Cosmological Attractor to General Relativity.}
In \Sref{br-di}, we demonstrate that the $\Lambda$CDM model serves as an attractor in BD theory with certain potentials. Additionally, we highlight in this section that even within more intricate scalar-tensor theories, general relativity itself can emerge as an attractor.

Different conformal frames are not equivalent to each other in a cosmological
perspective. However, once the physical conformal frame is determined, different
conformal frames are equivalent mathematically. Usually, people regard the
Jordan frame as the physical one, as baryons follow the geodesic and we can
directly compare the results with observations. In this case, the Einstein frame
is a mathematical treatment that can sometimes be used to avoid complicated
calculations in the original Jordan frame, e.g. Damour's famous work \cite{Damour:1993id,Damour:1996ke}.

According to the transformation \eref{CT}, it is easy to notice that
\begin{equation}
	T^{*\mathrm{m}}_{\mu\nu} =A^2T^\mathrm{m}_{\mu\nu},\quad T^\mu_{*\mathrm{m}\nu}=A^4T^{\mu}_{\mathrm{m}\nu},\quad
	T^{\mu\nu}_{*\mathrm{m}}=A^6T_\mathrm{m}^{\mu\nu},
\end{equation}
and the transformation of the trace is
\begin{equation}
	T_\mathrm{m}^*=A^4T_\mathrm{m}.
\end{equation}
One also defines $t_*$ and $a_*$ as
\begin{eqnarray}
	t_* \equiv\int A^{-1}(t)\rmd t,\\
	a_* \equiv A^{-1}a,
\end{eqnarray}
so that the line element is still in the form of the FLRW metric
\begin{equation}\label{RW}
	\rmd s_*^2=-\rmd t^2_*+a_*^2(t_*)\left(\frac{\rmd r^2}{1-Kr^2}+r^2\rmd\Omega^2_2\right).
\end{equation}

Substituting the metric \eref{RW} into equations of motion \eref{cor1} and
\eref{cor2}, we obtain
\begin{eqnarray}
	3H_*^2=\kappa_*(\rho_\mathrm{m}^*+\rho_\phi),\label{F1}\\
	-2\dot H_*=\kappa_*(\rho_\mathrm{m}^*+p_\mathrm{m}^*+\rho_\phi+p_\phi),\label{F2}\\
	\ddot\phi+3H_*\dot\phi+\frac{\rmd V}{\rmd\phi}=-\alpha(\rho_\mathrm{m}^*-3p_\mathrm{m}^*),
\end{eqnarray}
where the overdot represents the derivative with respect to the cosmic time
$t_*$, and $H_*\equiv\dot a_*/a_*$ is the Hubble parameter in the Einstein
frame. The Bianchi identity
\begin{equation}
	\rmd(\rho_\mathrm{m}^* a_*^3)+p_\mathrm{m}^*\rmd a_*^3=(\rho^*_\mathrm{m}-3p^*_\mathrm{m})a_*^3\rmd\ln A(\phi)
\end{equation}
tells us that the Einstein frame scaling laws are $\rho_\mathrm{r}\propto a_*^{-4}$ for
the radiation and $\rho_\mathrm{d}\propto A(\phi)a_*^{-3}$ for the dust.

Defining the $e$-folding number $N_*\equiv\ln a_*$ in the Einstein frame and
introducing parameters \cite{Damour:1992kf}
\begin{equation}
	w(N_*,\phi)\equiv\frac{p_\mathrm{m}^*}{\rho_\mathrm{m}^*},\quad v(N_*,\phi)\equiv\frac{V}{\rho_\mathrm{m}^*},
\end{equation}
the equation of motion of the scalar field becomes
\begin{equation}\label{phiparti}
	\frac{4(1+v)}{6-\kappa_*\phi'^2}\phi''+(1-w+2v) \phi'
	+\frac{2}{\kappa_*}\left[(1-3w)\alpha+v\frac{\rmd\ln V}{\rmd\phi}\right]=0,
\end{equation}
where the prime denotes the derivative with respect to $N_*$.

\begin{figure}[h]
    \centering
    \includegraphics[width=0.75\textwidth]{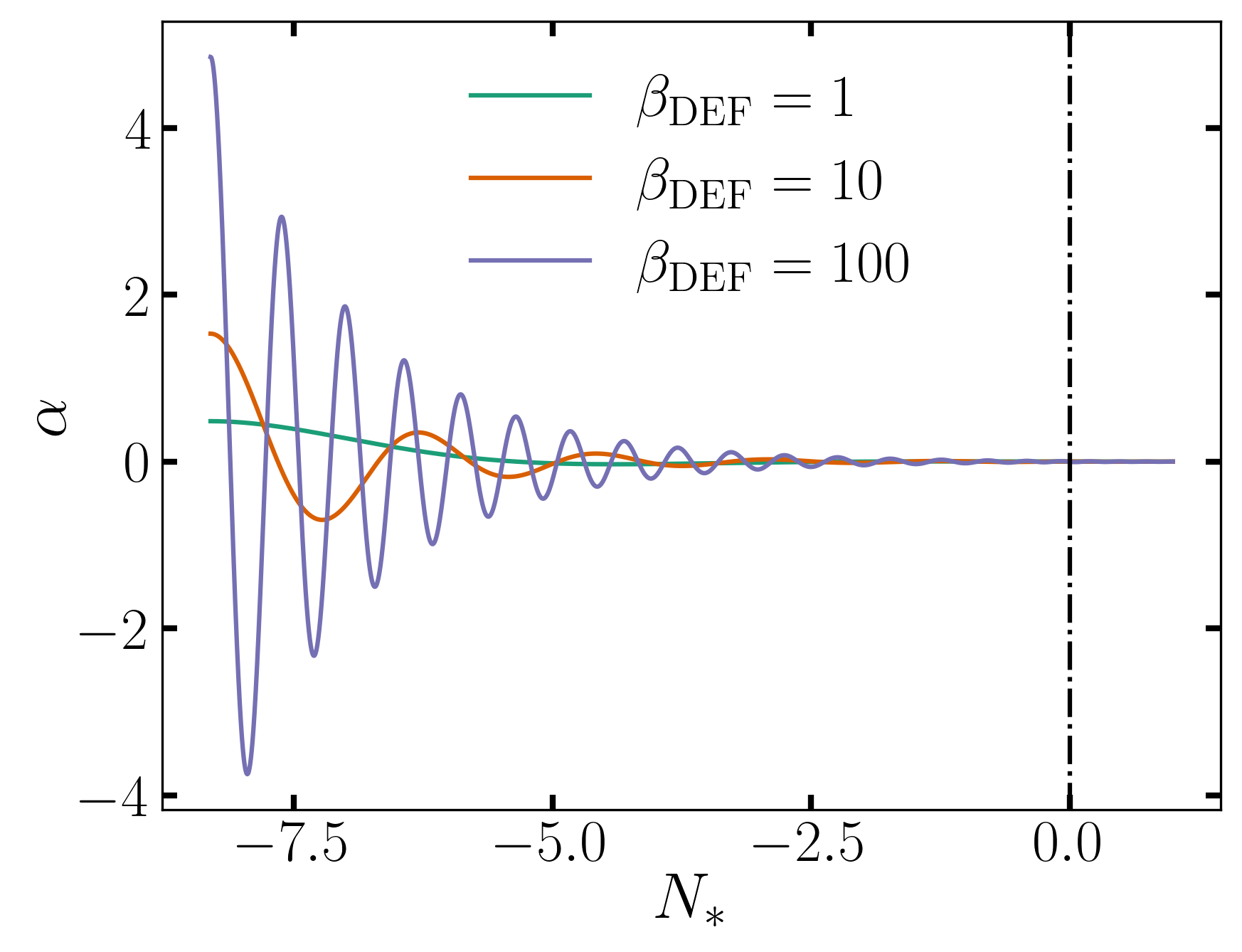}
    \caption{The evolution of the coupling $\alpha$ between matter and scalar field with different $\beta_\mathrm{DEF}$ in DEF theory. The initial conditions are $\phi'=0$ and $\phi$ saturating the constraint of speed-up factor during nucleosynthesis \cite{Uzan:2010pm}. The present coupling $\alpha_0=\alpha|_{N_*=0}$ is negligibly small.}
    \label{f:DEF}
\end{figure}

Each scalar-tensor theory is specified by a particular choice for $\alpha(\phi)$
and $V(\phi)$. For example, a constant $\alpha(\phi)=\alpha_\mathrm{BD}$ and
$V(\phi)=0$ select the traditional BD theory (\ref{BDA}). The mechanism of
attraction towards general relativity can be illustrated by the simplest case
where $\alpha(\phi)=\beta_\mathrm{DEF}\phi$, known as the Damour-Esposito-Far\`ese
(DEF) theory \cite{Damour:1993hw}, where in cosmology one usually consider the
case $\beta_\mathrm{DEF}>0$. Choosing $\kappa_*=2$ and $V=0$, \Eref{phiparti} takes the form of the equation of motion of a particle with a
velocity-dependent mass $m(\phi')=2/(3-\phi'^2)$ in a parabolic potential
$V_*(\phi)=(1-3w)\beta_\mathrm{DEF}\phi^2/2$, and subjecting to a damping force
proportional to $1-w$ \cite{Damour:1993id}. Notice that the potential $V_*$ vanishes
in the radiation-dominated epoch no matter what value $\alpha$ is. The exact
solution to the initial condition $(\phi,\phi')|_{N_*=0}=(\phi_\rmi,\phi_\rmi')$ for
the radiation epoch is \cite{Damour:1993id}
\begin{equation}
	\fl\phi(N_*) =\phi_\rmi-\sqrt{3}\ln\left(\frac{K_\rmi\rme^{-N_*}
	+\sqrt{1+K_\rmi^2\rme^{-2N_*}}}{K_\rmi+\sqrt{1+K_\rmi^2}}\right),\quad
	K_\rmi=\frac{\phi_\rmi'}{\sqrt{3-\phi_\rmi'^2}}.
\end{equation}
The duration of the radiation-dominated epoch $\Delta N_*^\mathrm{rad}\sim20$ is
long enough to slow down the particle even if it is ultra-relativistic as long
as $\phi'_i\lesssim\sqrt3$ \cite{Anderson:2016aoi,Coc:2006rt}. Then the particle is stationary at the beginning of
the matter-dominated epoch. Hence, it is easy to realize that at late times the field
$\phi$ will settle down at the minimum of the potential $V_*$ with $\phi=0$,
where $\alpha=\beta_\mathrm{DEF}\phi=0$ and the theory flows towards general relativity as the Universe evolves. \Fref{f:DEF} illustrates the evolution of $\alpha$ as the Universe evolves, revealing that the coupling between matter and scalar fields is currently negligibly small.

Another interesting feature of the DEF theory is that when the gravitational
field gets stronger, the compact star may scalarize spontaneously and become a
scalarized star if the parameter $\beta_\mathrm{DEF}\lesssim-4$ \cite{Damour:1993hw,Doneva:2022ewd}. Spontaneous scalarization and general
relativity attractor share different parameter spaces in the DEF theory and some
generalized theories \cite{Damour:1992kf,Anderson:2016aoi,Anderson:2017phb}. Adding another
extra coupling between the scalar field and the Gauss-Bonnet term is a feasible
solution to exhibit spontaneous scalarization in the strong-field regime and the
spacetime behaves like general relativity at infinity as a cosmological
attractor \cite{Antoniou:2020nax,Erices:2022bws}.

In order to explain the accelerated expansion, the potential $V(\phi)$ cannot be zero.
If one is interested in a scalar field with energy density of the same order as
that of matter today, then the term $\rmd\ln V/\rmd\phi$ in \Eref{phiparti} turns out to
be subdominant with respect to $(1-3w)\alpha$ during the radiation-domination
epoch and most of the matter-domination epoch. At late times, the scalar
potential term starts to dominate and the gravity is approximately described by
general relativity as $\alpha\to0$. By choosing the coupling function as
\begin{equation}
	\alpha(\phi)=-Be^{-\beta\phi}
\end{equation}
and the Ratra-Peebles potential \eref{RPV} for the scalar field, one obtains an
approximate solution during the radiation and matter domination
\cite{Bartolo:1999sq}
\begin{equation}
	\phi(a_*)\simeq\frac{1}{\beta}\ln\left[\beta B\ln
	\left(\frac23+\frac{a_*}{a_*^{\mathrm{eq}}}\right)+C\right],
\end{equation}
where $a_*^\mathrm{eq}$ is the value of the scale factor at the equivalence epoch (the
cosmic time when $\rho_\mathrm{r}=\rho_\mathrm{d}$) and $C$ is a constant. This solution is an
attractor in phase space and behaves as a subdominant contribution to the energy
density that starts increasing during the matter-dominated era and eventually comes to dominate in the
present era. The Doppler peaks are shifted toward higher multipoles and their
height is changed with respect to minimally coupled models, providing a
signature to look for in future CMB experiments \cite{Bartolo:1999sq}.

\subsection{Recapitulation and Possible Extension}

A general action describing various couplings to the scalar field for different components of the Universe in the Einstein frame is given by
\begin{equation}\label{EGA}
	\fl S[g^*_{\mu\nu},\phi,\Psi_\mathrm{m}^{(i)}]=\int
	\rmd^4x\sqrt{-g_*}\left[\frac{R_*}{2\kappa_*}-\frac12g_*^{\mu\nu}
	\partial_\mu\phi \partial_\nu\phi-V(\phi)\right]+S_\mathrm{m}[g_{\mu\nu}^{(i)},\Psi_\mathrm{m}^{(i)}]
\end{equation}
where $\Psi_m^{(i)}$ is the matter field coupled to the Jordan frame metric
$g_{\mu\nu}^{(i)}$, related to the Einstein frame metric $g^*_{\mu\nu}$ via
\begin{equation}
	g_{\mu\nu}^{(i)}=A^2_i(\phi)g^*_{\mu\nu}.
\end{equation}

The action \eref{EGA} reduces to the general relativity if $A_{i}=1$, which
means that each component of the Universe does not couple to the metric in
the Einstein frame. In that case, metrics in the Jordan frame and the Einstein
frame are the same. The quintessence model, in the framework of general
relativity, relies on a self-interacting potential of the scalar field and a
violation of the strong energy condition to provide necessary cosmic
acceleration. During a slow-roll phase of the massive scalar field when its
kinetic energy is much less than its self-interacting one, negative pressures
are achieved, producing the desired cosmic acceleration. The potential, as the
only degree of freedom in the quintessence model, can be quite sophisticated and
has to be well shaped to reproduce the cosmic evolution or to exhibit
interesting tracking properties. The coincidence problem is somehow moved to the specific choice of both an appropriate shape and energy scale for the
self-interaction potential. Most of the potentials used in quintessence models
come out of an effective theory of high-energy physics, leaving a glimpse into new physics. However, deriving such quintessence models (and also
the k-essence extension models) from high-energy physics in a self-consistent
framework is a challenging task, as combining all the cosmological,
gravitational, and particle physics aspects raises many technical difficulties.

The quintessence scalar field minimally couples to gravitation and does not
couple directly to ordinary matter. Hence, quintessence only modifies the
background cosmic expansion and all the observations of gravity theory without
the quintessence component remain the same. If matter fields couple to both
scalar field and gravitation directly, the new ``fifth" interaction yields a
violation of the strong equivalence principle \cite{Joyce:2016vqv}. The
non-minimal coupling changes physical coupling constants or inertial masses and
as a consequence, modifies the way energies have been weighted in the cosmic
history. In the case where the scalar field directly affects the couplings to
the gravitational field and therefore makes the gravitational constant
vary, one gets the scalar-tensor theories of gravity,
\begin{equation}
	\fl S[g^*_{\mu\nu},\phi,\Psi_\mathrm{m}] =\int \rmd^4x\sqrt{-g_*}
	\left[\frac{R_*}{2\kappa_*}-\frac12g_*^{\mu\nu}\partial_\mu
	\phi\partial_\nu\phi-V(\phi)\right]
	+S_\mathrm{m}[A^2(\phi)g_{\mu\nu}^*,\Psi_\mathrm{m}].
\end{equation}
If we consider that about 70\% of the missing energy is the whole contribution of such
a non-minimally coupled scalar field, it might be difficult to match the
present tests of gravity without assuming the non-minimal couplings to be
extremely weak. In fact, the coupled quintessence model corresponds to the case
$A_\mathrm{b}=1$ for baryons while $A_{\rm dm}$ for dark matter varies with $\phi$, which
reveals a non-universal coupling. This non-universal coupling solves the extremely small coupling problem, but the weak equivalent principle is severely
violated. As for the chameleon cosmology, however, both baryons and dark matter are
non-minimally coupled to the quintessence field as $A_\mathrm{b}=A_{\rm
dm}=e^{\alpha\phi}$, with the mass of the scalar field being environmentally
dependent. This offers the possibility of accounting for the present bounds of
local tests of the strong and weak equivalent principles on the Earth and in the
Solar system by the larger mass of the scalar field in a denser neighborhood
than in the cosmological low-density limit. In order to do so, a specific
self-interaction potential and a consequent violation of the strong energy
condition have to be re-introduced. Another alternative model proposes that both
$A_\mathrm{b}$ and $A_{\rm dm}$ depends on the scalar field, but not identical
\cite{Fuzfa:2007sv}. The self-interacting potential $V(\phi)$ is unnecessary in
this model if two couplings are both assumed to be of order unity, and the
cosmic acceleration is a generic prediction of the competition between different
non-minimal couplings. Moreover, the attractor to which gravitation is driven
depends on the ratio between the energy densities of ordinary matter and dark
matter.

The most general 4-dimensional scalar-tensor theories having second-order
equations of motion are the Horndeski theory
\cite{Nicolis:2008in,Deffayet:2009mn}
\begin{equation}\label{Horn}
	S_\mathrm{H}[g_{\mu\nu},\phi,\Psi_\mathrm{m}] =\sum_{i=2}^5\int \rmd^4x\sqrt{-g}
	\mathcal L_i+S_\mathrm{m}[g_{\mu\nu},\Psi_\mathrm{m}],
\end{equation}
where the Lagrangian densities are
\numparts
\begin{eqnarray}
	\fl\mathcal L_2 =G_2(\phi,X),\\
	\fl\mathcal L_3 =G_3(\phi,X)\square\phi,\\
	\fl\mathcal L_4 =G_4(\phi,X)R+G_{4X}\big[ (\square\phi)^2-\nabla_\mu
	\nabla_\nu\phi\nabla^\mu \nabla^\nu\phi\big],\\
	\fl\mathcal L_5 =G_5(\phi,X) G_{\mu \nu} \nabla^\mu \nabla^\nu
	\phi\nonumber\\-\frac{G_{5X}}{6} \Big[(\square \phi)^3-3 \square
	\phi\nabla_\mu\nabla_\nu\phi \nabla^\mu\nabla^\nu\phi+2\nabla^\mu
	\nabla_\alpha \phi\nabla^\alpha\nabla_\beta
	\phi\nabla^\beta\nabla_\mu\phi\Big].
\end{eqnarray}
\endnumparts
The theory is termed as Horndeski theory because Kobayashi et al. \cite{Kobayashi:2011nu} found that the action \eref{Horn} is equivalent to the
one discovered by Horndeski in 1974 \cite{Horndeski:1974wa}. Here,
$G_i$ ($i=2,3,4,5$) are arbitrary functions of the scalar field $\phi$ and its
kinetic energy $X$, while $G_{i\phi}$ and $G_{iX}$ represent the derivatives of
functions $G_i$ with respect to $\phi$ and $X$, respectively.  Due to the
second-order property, there is no Ostrogradski instability \cite{Ostrogradsky:1850fid} associated with the
Hamiltonian unbounded from below.

In a flat FLRW background, the EOS parameter of the scalar field $\phi$ is
\begin{equation}
	w_\phi=-1+\frac{2(q_\mathrm{t}-M_\mathrm{Pl}^2)\dot
	H-D_6\ddot\phi+D_7\dot\phi}{\rho_\phi},
\end{equation}
where
\begin{equation}
	q_\mathrm{t}=2G_4-2\dot\phi^2G_{4X}+\dot\phi^2G_{5\phi}-H\dot\phi^3G_{5X},
\end{equation}
and the definition of $D_{6,7}$ and energy density $\rho_\phi$ can be found in
Ref.~\cite{Kase:2018aps}.  Since the evolution of $w_\phi$ is different
depending on dark energy models, it is possible to distinguish them from
observations, like SN Ia, CMB, BAO,  LSSs, and weak lensing.

The propagation speed squared of the tensor perturbations in the Horndeski
theories is \cite{Kobayashi:2011nu}
\begin{equation}
	c_\mathrm{t}^2=\frac{1}{q_\mathrm{t}} \Big(2G_4-\dot\phi^2G_{5\phi}-\dot\phi^2\ddot\phi G_{5X} \Big).
\end{equation}
The detection of gravitational waves by GW170817 \cite{LIGOScientific:2017vwq}
from a binary neutron star merger together with the short gamma-ray burst GRB
170817A \cite{Goldstein:2017mmi} constrained the propagation speed $c_\mathrm{t}$ of
gravitational waves to be \cite{LIGOScientific:2017zic}
\begin{equation}
	-3\times10^{-15}<c_\mathrm{t}-1<7\times10^{-16}.
\end{equation}
If any fine tuning among functions is forbidden, one has
$G_{4X}=G_{5\phi}=G_{5X}=0$, then the Horndeski theory is restricted to be of
the form \cite{Baker:2017hug,Creminelli:2017sry,Sakstein:2017xjx,Ezquiaga:2017ekz}
\begin{equation}\label{rH}
	\fl S[g_{\mu\nu},\phi,\Psi_\mathrm{m}]=\int
	\rmd^4x\sqrt{-g}\big[G_2(\phi,X)+G_3(\phi,X)\square
	\phi+G_4(\phi)R\big]+S_\mathrm{m}[g_{\mu\nu},\Psi_\mathrm{m}].
\end{equation}

Each dark energy model or scalar-tensor gravity mentioned above corresponds to
this reduced action \eref{rH} with specific functions $G_i$ ($i=1,2,3$). The
k-essence model \eref{KEA} is given by the choice
\begin{equation}
	G_2=P(\phi,X),\quad G_3=0,\quad G_4=\frac{M^2_\mathrm{Pl}}2,
\end{equation}
and the extended quintessence model \eref{EQAA} is acquired if 
\begin{equation}
	G_2=\zeta(\phi)X-U(\phi),\quad G_3=0,\quad G_4=\frac{M^2_\mathrm{Pl}}{2}F(\phi).
\end{equation}

There are other modified gravity theories dubbed Galileon theories
\cite{Nicolis:2008in,Deffayet:2009wt,Deffayet:2009mn} containing scalar
derivative self-interactions $G_3\neq0$. In the original Galileon theory \cite{Nicolis:2008in}, the field
equations of motion are invariant under the Galilean shift
$\partial_\mu\phi\to\partial_\mu\phi+b_\mu$ in the Minkowski spacetime. In a
curved spacetime, the Lagrangian of covariant Galileon theories
\cite{Deffayet:2009mn} is constructed to keep the equations of motion up to the
second order, while recovering the Galilean shift symmetry in the Minkowski
limit. For covariant Galileons, there exist self-accelerating de Sitter
attractors responsible for the late-time cosmic acceleration
\cite{Gannouji:2010au,DeFelice:2010pv,DeFelice:2010nf}. However, the covariant
Galileons with quartic and quintic Lagrangians do not pass the test of tensor
perturbation without fine-tuning parameters.

The generally considered function of $G_3$ is the form of cubic Galileon
$X\square\phi$, which arises in the Dvali-Gabadadze-Porrati braneworld model due
to the mixture between longitudinal and transverse gravitons \cite{Dvali:2000hr}
and also in the Dirac-Born-Infeld decoupling theory with bulk Lovelock
invariants \cite{deRham:2010eu}. This derivative self-interaction can suppress the propagation of fifth forces in
local regions of the Universe with the Vainshtein mechanism
\cite{Vainshtein:1972sx}, while modifying the gravitational interaction at
cosmological distances \cite{Babichev:2009us,Babichev:2010jd,Burrage:2010rs,Kase:2013uja}. 

There are two kinds of solutions that displace
the present cosmic acceleration in minimally coupled cubic Galileon theories
($G_4=M_\mathrm{Pl}^2/2$). One is the solution with a constant $X$ in the absence
of potential \cite{DeFelice:2010pv,DeFelice:2010nf}, but it is in tension with
the observational data of redshift-space distortions, weak lensing, and
integrated Sachs-Wolfe-galaxy cross-correlations \cite{Renk:2017rzu}. The other
is the solution in cubic Galileon with a linear potential $V\propto\phi$ or other specific form \cite{Ali:2012cv}, and it
is the potential that drives the late-time accelerated expansion \cite{Deffayet:2009wt}. 
In recent years, the Galileon ghost condensate with $G_2=X+c_2X^2$ has received more attention for alleviating the problem of tracker solutions found in other cubic Galileon theories and improving the compatibility with the Planck CMB data \cite{Peirone:2019aua}. 
Furthermore, the cosmology
of non-minimally coupled cubic Galileons has also been studied
\cite{Kase:2015zva,DeFelice:2010nf}, including constraints on the coupling from
Solar system and binary pulsar tests \cite{Babichev:2011iz,Tsujikawa:2019pih,deRham:2012fw,Shao:2020fka}.

It is conceivable that the evolution of the Universe within Horndeski theory undergoes a diverse and dynamic process. Various phenomena such as late-time tracking, scaling, quintessence, and phantom behaviors are within the realm of possibility, which offer promising avenues for addressing the fine-tuning and coincidence problems. However, the intricate equations of motion arising from these model prevent simple applications of dynamical systems theories and extremely complicated analysis are required, often dealing with non-compact high-dimensional autonomous systems.

\section{A Brief Outlook}
We have presented a concise overview of two distinct physical mechanisms---dark energy and modified gravity---to elucidate the accelerated expansion of the Universe. The technique of dynamical system is applied to typical models---quintessence, coupled quintessence, and extended quintessence---to qualitatively describe the evolution of the Universe and the emergence of an accelerated expansion stage. The primary focus of the discussion revolves around constant couplings and exponential potentials, while more specific and sophisticated models are left to be explored in the original literature.

By assuming that the vacuum energy is either negligible or does not contribute to the gravitational field equation, modern study of cosmological constant problem on the classical level inquiries into identifying mechanisms responsible for the current acceleration of the Universe, so as to avoid the original fine-tuning problem in $\Lambda$CDM model. Nevertheless, for theories like quintessence to accurately portray the present state of the Universe, characterized by roughly equal proportions of matter and dark energy, the initial conditions of the Universe often need to be finely tuned to an extremely narrow range, thus presenting another instance of fine-tuning. A solution to the so-called coincidence problem would be to establish that the state where $\Omega_\mathrm{m}\simeq0.3$ and $\Omega_\mathrm{de}\simeq0.7$ is indeed the ultimate fate of the Universe, such as the coupled quintessence model. Two additional strategies that potentially alleviate the coincidence problem include expanding the range of initial conditions to accommodate existing observations of the Universe's evolution and permitting the Universe to evolve at a gradual pace towards a state vastly different from its current configuration, with tracking solutions and some interacting dark energy model being corresponding examples. Unlike the other two issues, the Hubble tension and more tensions observed in cosmological observations represent genuine problems that demand resolution but continue to harbor uncertainty. The interacting dark energy and chameleon dark energy discussed in this review have opportunity to solve these tensions. While these tensions are receiving increased attention, the scope of this review does not delve deeply into the early Universe, and more information can be found in comprehensive review papers such as Refs.~\cite{DiValentino:2021izs,Abdalla:2022yfr,Schoneberg:2021qvd}.

At present, dark energy models are not particularly favored over the
$\Lambda$CDM model from the current observational data, though they are better equipped to deal with various challenges. Furthermore, there is no concrete observational signature for non-minimally coupled theories, despite that they usually have much richer properties in dynamics. 
In the coming decades, LSS and CMB surveys will possess the capability to probe dynamical dark energy and deviations from general relativity with greater precision \cite{LSSTScience:2009jmu,EUCLID:2011zbd,SimonsObservatory:2018koc,Abazajian:2019eic}. Additionally, new methods of observing cosmology like gamma-ray burst detection \cite{Bozzo:2024hsu,Adil:2024miw} and gravitational-wave observation \cite{Punturo:2010zz,Reitze:2019iox} will contribute to the discourse. These advancements hopefully will either rule out some dark energy and modified gravity models or yield a breakthrough in cosmology and  physics by confirming new fundamental ingredients.

Numerous models exist to elucidate the current accelerated expansion of the Universe through the employment of scalar fields; however, we are unable to discuss all of them due to the limitation of space. 
Certainly, there are models that explain the expansion in terms of fluids or other fields beyond scalar fields, while some specific theoretical models also propose alternatives to the solutions discussed in this review \cite{Li:2011sd}.
We encourage the audience to explore additional reviews and specialized literature for supplementary insights, both within and beyond the reference of this review. The references we quote absolutely do not represent all the research in the field, and we regret any oversight of significant literature.

\section*{Acknowledgement}

We thank Rui Xu for discussions. This work was supported by the National Natural
Science Foundation of China (11991053), the Beijing Natural Science
Foundation (1242018), the National SKA Program of China (2020SKA0120300), the
Max Planck Partner Group Program funded by the Max Planck Society, and the
High-Performance Computing Platform of Peking University.

\section*{References}
\bibliographystyle{iopart-num-long}
\bibliography{refs}

\end{document}